\documentclass[]{article}
\usepackage{longtable}
\usepackage{amsmath,amssymb}
\usepackage{lmodern}
\usepackage[table]{xcolor}
\usepackage{colortbl}
\usepackage[margin=1in]{geometry}
\usepackage{color}
\usepackage{fancyvrb}
\usepackage{graphicx}
\usepackage{booktabs}
\usepackage{float}
\usepackage[hypertexnames=false]{hyperref}
\title{Power Studies For Goodness-of-Fit and \newline Two-Sample Methods For Univariate Data}
\author{Wolfgang Rolke}
\date{
  University of Puerto Rico - Mayaguez
	\newline
	\texttt{wolfgang.rolke@upr.edu}
  \newline
  \today
}

\begin{document}
\maketitle

\begin{abstract}
We report a large simulation study of the power of goodness-of-fit and nonparametric two-sample tests for univariate data, both continuous and discrete. In most of the 120 goodness-of-fit and 100 two-sample case studies the null hypothesis and the alternative agree in three of the first four moments and differ only in the mean, the variance, the skewness or the kurtosis. No single method has good power throughout: a method that is best in one case study may fail badly in another. We therefore propose for each problem a small selection of methods such that in every case study at least one of them has a power within 90\% of the best, checked by leave-one-out cross-validation. For continuous data Zhang's tests have the highest mean power, but much less so when parameters are estimated. For discrete data the data-driven smooth tests and the Berk-Jones test are best in the goodness-of-fit problem, the Epps-Singleton and HHG tests in the two-sample problem. Hybrid tests, which compare the data with a Monte Carlo sample from the null hypothesis, have lower power than goodness-of-fit tests. The methods, all results and the scripts are in the R packages \emph{Rgof} and \emph{R2sample}.
\end{abstract}

\section{Introduction}\label{introduction}

Both the goodness-of-fit (gof) and the nonparametric two-sample problem
have histories going back a century, with many contributions by some of
the most eminent statisticians. In the goodness-of-fit problem we have a
sample \((x_1,..,x_n)\) drawn from some probability distribution \(F\),
possibly with unknown parameters, and we wish to test \(H_0:X\sim F\).
In the two-sample problem we also have a second sample \((y_1,..,y_m)\)
from some distribution \(G\), and here we want to test \(H_0:F=G\), that
is we want to test whether the two data sets were generated by the same
(unspecified) distribution.

The literature on both of these problems is vast and steadily growing.
Detailed discussions can be found in \cite{agostini1986},
\cite{thas2010}, \cite{raynor2009}. For an introduction to
Statistics and hypothesis testing in general see \cite{casella2002} or \cite{bickel2015}.

The power studies in this article were carried out using \textbf{R}
programs in the packages \emph{Rgof} and \emph{R2sample}, available from
the CRAN website. Some tests such as the Kolmogorov-Smirnov test are
already implemented for both problems in base \textbf{R}. Many others
can be run through various packages, for example the Anderson-Darling
goodness-of-fit test is available in the \textbf{R} package
\emph{ADGofTest} \cite{ad2011}. There are also packages that allow the
user to run several tests, for example the \emph{twosamples} \cite{dowd2022}
package. However, there are no packages that bring together as many
tests as \emph{Rgof} and \emph{R2sample}. Also

\begin{itemize}
\item
  almost all the methods are implemented for both continuous and discrete
  data. Discrete data includes the case of histogram (aka discretized or binned) data.
\item
  the methods are implemented using both \emph{Rcpp} \cite{rcpp2024} and
  parallel programming.
\item
  some of the methods allow for data with weights.
\item
  the routines allow for a random sample size, assumed to come from a
  Poisson distribution.
\item
  in the two-sample problem some methods make use of large-sample
  formulas, therefore allowing for very large data sets.
\item
  there are routines that allow the user to combine several tests and
  find a corrected p value.
\item
  the routines can also use any other user-defined tests.
\item
  the packages include routines to easily run a large number of power studies and thereby compare a new method's performance to those included in the packages.
\end{itemize}

Including tests for discrete data is useful in two ways: Of course
discrete data is of interest in its own right, and there are few
implementations in \emph{R} at this time. It also makes it possible to apply
the tests to very large continuous data sets via discretization. While a
test for a continuous data set with (say) 100,000 observations
can be done in a matter of a few minutes, for larger data sets the
calculations will be quite time consuming. Binning the data and then
running the corresponding discrete tests however is quite fast.

There are also situations where the underlying distribution is
continuous but the data is collected in binned form. This is for example
often the case for data from high energy physics experiments and from
astronomy because of finite detector resolution. In this situation the
theoretical distribution is continuous but the data is discrete.

For the tests in the two-sample problem p values are found via the
permutation method. If the data sets are large, for some of the
tests the p values can be found via large sample approximations. In the
goodness-of-fit case p values are found via simulation (aka parametric bootstrap). While
large sample approximations are known for some methods such as
Kolmogorov-Smirnov and Anderson-Darling, there are no known large sample
theories for most of the other tests. Moreover, in the more common
situation where the distribution under the null hypothesis depends on
parameters, which have to be estimated from the data, even those tests
no longer have known large sample theories and one is forced to use
simulation to find p values. The exceptions are the chi-square tests, which use the usual chi-square approximation.

Our aim is to use these tests for their original purpose: a researcher wishes to check whether a probability model, which a subsequent analysis will rely on, is, in the words of George Box, ``good enough''. Both problems are considered for continuous and for discrete data; discrete data, which includes histogram (binned) data, consist of a set of values and the counts of the observations at these values. In the goodness-of-fit problem the null hypothesis can be simple, with $F$ fully specified, or composite, with parameters that have to be estimated from the data.

This paper is a companion to a study of the same kind for multivariate data with the packages \emph{MDgof} and \emph{MD2sample}. The two studies use the same methodology to summarize the results.

\section{Goodness-of-fit/two-sample hybrid tests}

There is another (quite popular) way to carry out a goodness-of-fit test. In these tests one generates a second data set $y$ under the null hypothesis, using a parametric bootstrap approach if parameters are estimated. Then one carries out a two-sample test. In \emph{Rgof} this is implemented in the routine \emph{hybrid\textunderscore test}, which uses the methods of \emph{R2sample}.

There are a number of reasons one might do this:
\begin{itemize}
\item
One can now employ any two-sample method.
\item
The actual distribution function $F$ might not be calculable, for example because it requires numerical integration of a density that is only known up to a constant.
\end{itemize}

This approach does raise an additional issue. If generating data from $F$ is fast, how large a sample should be generated, relative to the sample size of the real data? Generally one would expect the power to increase with a larger data set.

In this paper we include this hybrid method with two sample sizes for the MC data set, one equal to the size of the real data (nMC=1) and another five times as large (nMC=5).

\section{Highlights of the Results}\label{sec:highlights}

Detailed results of our studies can be found later, here are some highlights:

\begin{itemize}
\item
  As one would expect, no single method is uniformly better than all the others. In most case studies in which the best method has a power of about $80\%$ there are other methods with a power below $30\%$.
\item
  For continuous data and no parameter estimation the tests of Zhang, ZC and ZA, have by far the highest mean power, in both the goodness-of-fit and the two-sample problem. In the two-sample problem each of them is within $90\%$ of the best method in two thirds of the case studies, and in the case studies in which the two distributions differ only in their variance in $95\%$ of them. In the two-sample problem the Epps-Singleton test (mean power $67\%$) comes next, followed by the HHG test and Zhang's ZK (both $63\%$).
\item
  When parameters are estimated the picture changes. Watson's test and the Anderson-Darling test then have the highest mean power, and the Zhang tests lose most of their advantage.
\item
  For discrete data the Zhang tests are not available in the goodness-of-fit problem, and the discrete version of ZA in the two-sample problem has low power. In the goodness-of-fit problem the data-driven smooth tests NS-B and NS-A and the Berk-Jones test have the highest mean power ($71\%$, $68\%$ and $66\%$), followed by Kuiper, Anderson-Darling and Watson (about $60\%$). In the two-sample problem the Epps-Singleton and the HHG tests (both $83\%$) are far ahead of all the other methods.
\item
  Chi-square tests with a small number of bins (10) are part of the best selection for continuous data in the goodness-of-fit problem. With the new methods they are no longer needed for discrete data or in the two-sample problem.
\item
  Hybrid methods require a larger MC data set than the real data in order to be competitive, but even with an MC data set five times the size of the data they have lower power than the goodness-of-fit tests. So as long as it is possible to do a classic goodness-of-fit test, these are preferred.
\item
  For many case studies the best method for the discretized data has a higher power than the best method for the continuous data, because of the strong discrete versions of the smooth, Berk-Jones, Epps-Singleton and HHG tests.
\end{itemize}

Based on the results of these simulation studies our recommendation for the best methods are:

\begin{itemize}
\item
  Goodness-of-fit, continuous data, no parameter estimation: Zhang's ZC and ZA, Watson, Wasserstein, Anderson-Darling (or Cram\'er-von Mises) and a chi-square test with a small number of bins.
\item
  Goodness-of-fit, continuous data, with parameter estimation: Watson, Zhang's ZC and Wasserstein, possibly together with Anderson-Darling.
\item
  Goodness-of-fit, discrete data: the smooth tests NS-B and NS-A, Berk-Jones, Anderson-Darling, Wasserstein and Watson (or Kuiper).
\item
  Two-sample, continuous data: Zhang's ZC (or ZA), Epps-Singleton, HHG and Anderson-Darling (or Cram\'er-von Mises or Wasserstein).
\item
  Two-sample, discrete data: HHG, Anderson-Darling and Kuiper, possibly together with Epps-Singleton.
\item
  Hybrid tests only if the cdf under the null hypothesis can not be calculated, with an MC data set much larger than the data and the Zhang tests ZC and ZA as well as the HHG and Epps-Singleton tests.
\end{itemize}

\section{The Methods}\label{the-methods}

In the following we list the methods included in the packages. Most are
well known and have been in use for a long time. For their details see
the references. We use the following notations:

Goodness-of-fit problem, continuous data: we denote the cumulative distribution function (cdf) by
\(F\), the empirical distribution function (edf) by \(\widehat{F}\). The sample size is $n$ and the ordered data set is $x_1,..,x_n$.

Goodness-of-fit problem, discrete data: There are $k$ possible values $v_1,..,v_k$ that can be observed, the cdf is \(F\), the edf \(\widehat{F}\). The sample size is $n$ and the counts are $x_1,..,x_k$.

Two-sample problem, continuous data: There are data sets $x_1,..,x_n$ and $y_1,..,y_m$. The combined data set is denoted by $z_1,..,z_{n+m}$. These have edf's
 \(\widehat{F}\),  \(\widehat{G}\) and  \(\widehat{H}\), respectively.

Two-sample problem, discrete data: There are $k$ possible values $v_1,..,v_k$ that can be observed. The two data sets have counts $x_1,..,x_k$ and $y_1,..,y_k$. The combined data set has counts $z_i=x_i+y_i$. These have edf's \(\widehat{F}\),  \(\widehat{G}\) and  \(\widehat{H}\), respectively.

\subsection{Chi-Square Tests}

In the goodness-of-fit problem with continuous data \emph{Rgof} includes eight chi-square
tests, with either equal size (ES) or equal probability (EP) bins,
either a large (l=50) or a small (s=10) number of bins and with either
the Pearson (P) or the log-likelihood (L) formula. So the combination of
a large number of equal size bins and Pearson's chi-square formula is
denoted by ES-l-P, etc. In the two-sample problem \emph{R2sample} includes four tests with Pearson's formula, \emph{ES large}, \emph{ES small}, \emph{EP large} and \emph{EP small}.

In the case of discrete data the values are already given. They are combined into a large and a small number of classes, and the tests are denoted by l-P and s-P in the goodness-of-fit problem and by \emph{large} and \emph{small} in the two-sample problem.

In all cases neighboring bins with low counts are joined until all bins
have an expected count of at least 5 in the goodness-of-fit problem and a combined count of at least 5 in the two-sample problem. In all cases the p values are found using
the usual chi-square approximation.

If parameters have to be estimated, this is done via the user-provided
routine \(phat\). As long as the method of estimation used is consistent
and efficient and the expected counts are large enough the chi-square
statistic will have a chi-square distribution, with the degrees of freedom reduced by the number of estimated parameters, as shown by \cite{fisher1922}
and \cite{fisher1924}. This requires that the parameters are estimated from the binned data. By default (argument \emph{ChiUsePhat=FALSE}) the value provided by \emph{phat} is therefore used only as a starting point, and the parameters are estimated via the method of minimum chi-square. This
method has the desirable feature that if the null hypothesis is rejected
for this set of values, it will always be rejected for any other as
well. For a discussion of this estimation method see \cite{berkson1980}.

That a small number of bins usually leads to a higher power than a large number has been shown before, see for example \cite{rolke2021}.

\subsubsection{Goodness-of-Fit Problem}

Say the $i^{th}$ bin is $[a_i,a_{i+1}]$, $1\le i \le k$ and $O_i$ is the observed count in the $i^{th}$ bin. If $E_i=n\left[F(a_{i+1})-F(a_{i})\right]$ is the expected count in the $i^{th}$ bin, then the test statistics are given by

Pearson: $\sum_{i=1}^k \frac{(O_i-E_i)^2}{E_i} = \sum_{i=1}^k \frac{O_i^2}{E_i}-n$

log-likelihood: $2\sum_{i=1}^k \left[E_i-O_i+O_i\log \frac{O_i}{E_i}\right]$

Because a chi-square test is based on binned data, the formulas for discrete data are essentially the same.

If the large-sample approximation holds, then these test statistics will have a chi-square distribution with degrees of freedom df, where

\begin{itemize}
\item
  no parameter estimation, fixed sample size: df = number of bins - 1
\item
  no parameter estimation, random sample size: df = number of bins
\item
  with parameter estimation, fixed sample size: df = number of bins - 1 - number of estimated parameters
\item
  with parameter estimation, random sample size: df = number of bins - number of estimated parameters
\end{itemize}

\subsubsection{Two-Sample Problem}

Here we have the observed counts $O_i$ and $M_i$ for the two data sets in the $i^{th}$ bin. Let $N_1=\sum O_i$ and $N_2=\sum M_i$, then the chi-square test statistic is given by

$$\sum_{i=1}^{k} \frac{\left(\sqrt{N_2/N_1}\,O_i-\sqrt{N_1/N_2}\,M_i\right)^2}{O_i+M_i}$$

which is the same as Pearson's chi-square statistic for the $2\times k$ table of the counts, with the expected counts found by combining the two data sets. Under the null hypothesis this test statistic will have approximately a chi-square distribution with the degrees of freedom equal to the number of bins - 1.

\subsection{Kolmogorov-Smirnov (KS)}

This test is based on the largest absolute distance between \(F\) and
\(\widehat{F}\) in the goodness-of-fit problem and between
\(\widehat{F}\) and \(\widehat{G}\) in the two-sample problem. The tests
were first proposed in \cite{kolmogorov1933}, \cite{smirnov1939} and are among
the most widely used tests today. There is a known large sample
distribution of the test statistic in the two-sample problem, which is
used either if both sample sizes exceed 10000 or if the argument
\emph{UseLargeSample=TRUE} is set. In the goodness-of-fit case the large
sample theory is known only in the case of a fully specified
distribution under the null hypothesis. Because this is rarely of
interest the large sample approximation is not used.

\subsubsection{Goodness-of-Fit Problem}

\textbf{Theory}

$$\max\{|\widehat{F}(x)-F(x)|:x\in \mathbb{R}\}$$

\textbf{Continuous Data}

$$\max \left\{ \left|F(x_i)-\frac{i}{n}\right|,\left|\frac{i-1}{n}-F(x_i)\right|:i=1,..,n \right\}$$

\textbf{Discrete Data}

$$\max \left\{ |F(v_i)-\widehat{F}(v_i)|,|\widehat{F}(v_{i-1})-F(v_i)|:i=1,..,k \right\}$$

\subsubsection{Two-Sample Problem}

\textbf{Theory}

$$\max\{|\widehat{F}(x)-\widehat{G}(x)|:x\in \mathbb{R}\}$$

\textbf{Continuous Data}

$$\max \left\{ |\widehat{F}(z_i)-\widehat{G}(z_i)|:i=1,..,n+m \right\}$$

\textbf{Discrete Data}

$$\max \left\{ |\widehat{F}(v_i)-\widehat{G}(v_i)|:i=1,..,k \right\}$$

\subsection{Kuiper (K)}

This test is closely related to Kolmogorov-Smirnov, but it uses the sum
of the largest positive and negative differences as a test statistic. It
was first proposed in \cite{kuiper1960}. In \emph{R2sample} it is denoted by \emph{Kuiper}.

\subsubsection{Goodness-of-Fit Problem}

\textbf{Theory}

$$\max\{\widehat{F}(x)-F(x):x\in \mathbb{R}\}+\max\{F(x)-\widehat{F}(x):x\in \mathbb{R}\}$$

\textbf{Continuous Data}

$$\max \left\{ \frac{i}{n}-F(x_i):i=1,..,n \right\}+\max \left\{ F(x_i)-\frac{i-1}{n}:i=1,..,n \right\}$$

\textbf{Discrete Data}

$$\max \left\{ \widehat{F}(v_i)-F(v_i):i=1,..,k \right\}+\max \left\{ F(v_i)-\widehat{F}(v_{i-1}):i=1,..,k \right\}$$

\subsubsection{Two-Sample Problem}

\textbf{Theory}

$$\max\{\widehat{F}(x)-\widehat{G}(x):x\in \mathbb{R}\}-\min\{\widehat{F}(x)-\widehat{G}(x):x\in \mathbb{R}\}$$

\textbf{Continuous Data}

$$\max \left\{\widehat{F}(z_i)-\widehat{G}(z_i):i=1,..,n+m \right\}-\min \left\{ \widehat{F}(z_i)-\widehat{G}(z_i):i=1,..,n+m \right\}$$

\textbf{Discrete Data}

$$\max \left\{\widehat{F}(v_i)-\widehat{G}(v_i):i=1,..,k \right\}-\min \left\{ \widehat{F}(v_i)-\widehat{G}(v_i):i=1,..,k \right\}$$

\subsection{Cram\'er-von Mises (CvM)}

This test is based on the integrated squared differences. The goodness-of-fit version is discussed in \cite{cramer1928} and \cite{mises1928}. The
two-sample version was proposed in \cite{anderson1962}.

\subsubsection{Goodness-of-Fit Problem}

\textbf{Theory}

$$\int_{-\infty}^{\infty} (F(x)-\widehat{F}(x))^2dF(x)$$

\textbf{Continuous Data}

$$\frac1{12n}+\sum_{i=1}^n \left(\frac{2i-1}{2n}-F(x_i)\right)^2$$

\textbf{Discrete Data}

$$\frac1{12n}+n\sum_{i=1}^k \left(\widehat{F}(v_i)-F(v_i)\right)^2\left(F(v_i)-F(v_{i-1})\right)$$

\subsubsection{Two-sample Problem}

\textbf{Theory}

$$\int_{-\infty}^{\infty} (\widehat{F}(x)-\widehat{G}(x))^2d\widehat{H}(x)$$

\textbf{Continuous Data}

$$\frac{nm}{(n+m)^2}\sum_{i=1}^{n+m} \left(\widehat{F}(z_i)-\widehat{G}(z_i)\right)^2$$

\textbf{Discrete Data}

$$\frac{nm}{(n+m)^2}\sum_{i=1}^k z_i\left(\widehat{F}(v_i)-\widehat{G}(v_i)\right)^2$$

The Lehmann-Rosenblatt test (\cite{lehmann1951}, \cite{rosenblatt1952}), included in earlier versions of \emph{R2sample}, is equivalent to the two-sample Cram\'er-von Mises test and is no longer part of the package.

\subsection{Anderson-Darling (AD)}

This test is similar to the Cram\'er-von Mises test but with an integrand
that emphasizes the tails. It was first proposed in \cite{anderson1952}. The two-sample
version is discussed in \cite{pettitt1976}.

\subsubsection{Goodness-of-Fit Problem}

\textbf{Theory}

$$\int_{-\infty}^{\infty} \frac{(F(x)-\widehat{F}(x))^2}{F(x)(1-F(x))}dF(x)$$

\textbf{Continuous Data}

$$-n-\frac1{n}\sum_{i=1}^n (2i-1)\left(\log(F(x_i))+\log(1-F(x_{n+1-i}))\right)$$

\textbf{Discrete Data}

$$n\sum \frac{(F(v_i)-\widehat{F}(v_i))^2}{F(v_i)(1-F(v_i))}\left(F(v_i)-F(v_{i-1})\right)$$
where the sum runs over those $v_i$ with $0<F(v_i)<1$.

\subsubsection{Two-sample Problem}

\textbf{Theory}

$$\int_{-\infty}^{\infty} \frac{(\widehat{F}(x)-\widehat{G}(x))^2}{\widehat{H}(x)(1-\widehat{H}(x))}d\widehat{H}(x)$$

\textbf{Continuous Data}

$$nm\sum_{i=1}^{n+m-1} \frac{\left(\widehat{F}(z_i)-\widehat{G}(z_i)\right)^2}{i(n+m-i)}$$

\textbf{Discrete Data}

$$nm\sum_{i=1}^{k} w_i\left(\widehat{F}(v_i)-\widehat{G}(v_i)\right)^2$$
 where $w_i$ is the sum of $1/(j(n+m-j))$ over the observations $j$ of the combined data set at the value $v_i$.

\subsection{Wasserstein p=1 (Wassp1)}

A test based on the Wasserstein p=1 metric. It is based on a comparison
of quantiles. In the goodness-of-fit case these are the quantiles of the
data set and the quantiles of the cdf, and in the two-sample problem
they are the quantiles of the individual data sets and the quantiles of
the combined data set. If \(n=m\) the test statistic in the continuous
case takes a very simple form: \(\frac1n\sum_{i=1}^n |x_i-y_i|\), where $x$ and $y$ are ordered. In the
goodness-of-fit problem for continuous data the user has to supply a
function that calculates the inverse of the cdf under the null
hypothesis. For a discussion of the Wasserstein distance see \cite{wasserstein1969}.

\subsection{Zhang's tests (ZA, ZK and ZC)}

These tests were proposed in \cite{zhang2002} and \cite{zhang2006}. They
are variations of test statistics based on the likelihood ratio.

\subsubsection{Goodness-of-Fit Problem}

\textbf{Continuous Data}

$$
\begin{aligned}
&Z_K    = \max_{1\le i\le n} \left( (i-\tfrac12)\log\left\{\frac{i-\frac12}{nF(x_i)}\right\}+(n-i+\tfrac12)\log\left\{\frac{n-i+\frac12}{n[1-F(x_i)]}  \right\} \right) \\
&Z_A    = -\sum_{i=1}^n \left[\frac{\log [F(x_i)]}{n-i+\frac12}+\frac{\log [1-F(x_i)]}{i-\frac12}\right]\\
&Z_C    = \sum_{i=1}^n \left[\log\left\{\frac{F(x_i)^{-1}-1}{(n-\frac12)/(i-\frac34)-1}\right\} \right]^2
\end{aligned}
$$

\textbf{Discrete Data}

None of these tests have a version for discrete data. The reason is that they are built on the concept of ranks, and in the discrete data case there are simply too many ties.

\subsubsection{Two-sample Problem}

We define $R_{j}^x$ to be the rank of the $j^{th}$ order statistic of the x data set in the pooled data set, and $R_{j}^y$ accordingly. Let $N=n+m$ and $t_i=(i-\frac12)/N$ for $i=1,..,N$. Also let $z_i$ be the $i^{th}$ observation in the ordered pooled sample of $x$ and $y$.

\textbf{Continuous Data}

$$
\begin{aligned}
&Z_K=\max_{1\le i\le N}\left\{ n\left[\widehat{F}(z_i)\log\frac{\widehat{F}(z_i)}{t_i}+(1-\widehat{F}(z_i))\log\frac{1-\widehat{F}(z_i)}{1-t_i}\right] \right.+\\
&\left.m\left[\widehat{G}(z_i)\log\frac{\widehat{G}(z_i)}{t_i}+(1-\widehat{G}(z_i))\log\frac{1-\widehat{G}(z_i)}{1-t_i}\right] \right\}\\
&\\
&Z_A=-\sum_{i=1}^{N} \frac{n\,h\left(\widehat{F}(z_i)\right)+m\,h\left(\widehat{G}(z_i)\right)}{(i-\frac12)(N-i+\frac12)}, \text{ where } h(u)=u\log u+(1-u)\log(1-u)\\
&\\
&Z_C = \frac1n\sum_{i=1}^n \log\left(\frac{n}{i-\frac12}-1\right)\log\left(\frac{N}{R^x_i-\frac12}-1\right)+\\
&\phantom{Z_C = }\frac1m\sum_{i=1}^m \log\left(\frac{m}{i-\frac12}-1\right)\log\left(\frac{N}{R^y_i-\frac12}-1\right)
\end{aligned}
$$

\textbf{Discrete Data}

Only the $Z_A$ method has a discrete version, which uses a formula essentially the same as the continuous data formula.

\subsection{Watson's Test (W), Goodness-of-Fit Problem}

This test is closely related to the Cram\'er-von Mises test. It adjusts that
test statistic via a squared difference of the mean of
\(F(x_i)\) and 0.5, which makes it invariant to the choice of the origin, a property useful for circular data. It was proposed in \cite{watson1961}.

\subsection{Data-driven smooth tests (NS-B and NS-A), Goodness-of-Fit Problem, Discrete Data}

Smooth tests go back to \cite{neyman1937}, for a detailed discussion see \cite{raynor2009}. The data-driven version, in which the number of components is chosen from the data, was proposed in \cite{ledwina1994}. Let $p_j=F(v_j)-F(v_{j-1})$ and let $u_j=F(v_{j-1})+p_j/2$ be the mid-cdf of the $j^{th}$ value. The Legendre polynomials in $2u_j-1$ are made orthonormal with respect to the null probabilities $p_j$ (on the values with $p_j>0$), which gives functions $g_1,..,g_K$ with $\sum_j p_jg_l(v_j)g_r(v_j)=\delta_{lr}$, $\sum_j p_jg_l(v_j)=0$ and $K=\min(10, k-1)$. With

$$b_l=\frac1{\sqrt{n}}\sum_{j=1}^k x_jg_l(v_j) \text{ and } S_d=\sum_{l=1}^d b_l^2$$

the test statistic is $S_{d^*}$, where $d^*$ maximizes $S_d-d\log n$ (Schwarz's rule, NS-B) or $S_d-2d$ (Akaike's rule, NS-A). The Schwarz rule usually selects only the first one or two components and so is good at detecting differences in location or scale, whereas the AIC rule more readily includes higher components.

These tests are implemented in \emph{Rgof} only for discrete data. A continuous version based on $u_i=F(x_i)$ was included in our search for new methods, but it did not add anything to the methods already included for continuous data.

\subsection{Berk-Jones (BJ), Goodness-of-Fit Problem, Discrete Data}

This test was proposed in \cite{berk1979}. It is based on the largest likelihood ratio statistic for the binomial distribution of $n\widehat{F}(v_i)$:

$$BJ=n\max_i \left[\widehat{F}(v_i)\log\frac{\widehat{F}(v_i)}{F(v_i)}+(1-\widehat{F}(v_i))\log\frac{1-\widehat{F}(v_i)}{1-F(v_i)}\right]$$

where the maximum is over the $v_i$, $i<k$, with $0<F(v_i)<1$. It is closely related to Zhang's $Z_K$.

\subsection{Epps-Singleton (EppsS), Two-Sample Problem}

This test, proposed in \cite{epps1986}, compares the empirical characteristic functions of the two data sets at the two points $t_1=0.4/\hat{\sigma}$ and $t_2=0.8/\hat{\sigma}$, where $\hat{\sigma}$ is the semi-interquartile range of the combined data set. Let $g(z)=(\cos t_1z, \cos t_2z, \sin t_1z, \sin t_2z)$, let $\bar{g}_x$ and $\bar{g}_y$ be the means of $g$ over the x and the y data, and $\hat{\Omega}=\frac{n+m}{n}\hat{\Sigma}_x+\frac{n+m}{m}\hat{\Sigma}_y$, where $\hat{\Sigma}_x$ and $\hat{\Sigma}_y$ are the sample covariance matrices of $g$ over the two data sets. Then

$$W=(n+m)(\bar{g}_x-\bar{g}_y)'\hat{\Omega}^{+}(\bar{g}_x-\bar{g}_y)$$

where $\hat{\Omega}^{+}$ is the (pseudo) inverse of $\hat{\Omega}$. For discrete data the means and covariances are weighted by the counts. Under the null hypothesis $W$ has asymptotically a chi-square distribution with degrees of freedom equal to the rank of $\hat{\Omega}$ (usually 4), which is used for large data sets. See also \cite{goerg2009}.

\subsection{Heller-Heller-Kaufman-Brill-Gorfine (HHG), Two-Sample Problem}

This test is based on \cite{heller2016}. In its equipartition version, which is used here, the combined data set is divided into atoms: for continuous data 40 groups of consecutive order statistics of (nearly) equal size, for discrete data the distinct values. For each number of cells $m=2,..,10$ the statistic $S_m$ is the average of the likelihood ratio statistic of the $2\times m$ table of counts over all partitions of the atoms into $m$ contiguous cells, which can be found without going through all the partitions. The $S_m$ are combined with the MinP rule: each $S_m$ is turned into a p value using its distribution under random relabeling of the combined data set, and the test statistic is $-\log(\min_m p_m)$. Because it considers partitions with many cells the test has good power against alternatives where the densities differ in several places. For continuous data the test is distribution-free, and for large data sets the p value is found from the simulated null distribution without a permutation test.

\section{Case studies}\label{sec:cases}

The conclusions described in this paper regarding the powers of the various methods are based on numerous case studies. We have the following different types of studies:

\begin{enumerate}
\item
 goodness-of-fit or two-sample problem.
\item
 continuous or discrete data.
\item
 studies in which the null hypothesis and the alternative (or the two distributions in the two-sample problem) have the same mean, variance, skewness and kurtosis except for one of them, or studies in which several of these differ.
\item
 studies with or without parameter estimation (goodness-of-fit problem).
\end{enumerate}

The case studies are arranged by \emph{type}:

\begin{itemize}
\item \emph{mean}, \emph{variance}, \emph{skewness}, \emph{kurtosis}: there are 20 distributions, including uniform, beta, normal, gamma and exponential distributions and mixtures of these. For each of them and each of the four moments an alternative was constructed that has the same values as the null distribution for three of the first four moments and differs in the fourth. So for example in the case studies of type \emph{variance} the alternative has the same mean, skewness and kurtosis as the null distribution and differs only in the variance. These case studies are therefore quite difficult, any test has to detect a difference in only one of the moments. The alternatives are included in the packages in the data set \emph{funs\textunderscore list}.
\item \emph{mixed}: these are more traditional case studies, where the null distribution and the alternative differ in several moments, for example a uniform distribution and a uniform distribution with a bump, or an exponential and a Weibull distribution. In the goodness-of-fit problem there are 40 such case studies, 20 without and 20 with parameter estimation, and in the two-sample problem there are 20.
\end{itemize}

So in the goodness-of-fit problem there are $4\times 20+40=120$ case studies and in the two-sample problem $5\times 20=100$, each for both continuous and discrete data. The discrete case studies are discretized versions of the continuous ones, with 50 bins. Both \emph{Rgof} and \emph{R2sample} include vignettes that describe all the case studies in detail. Here we will only show a few select examples.

The alternative distribution generally depends on a parameter. In the results described in this paper there are always two of these. One is chosen so that the null hypothesis is actually true (and one can check that the methods achieve the nominal type I error) and a second one is chosen in such a way that for the selected sample size and a type I error of $\alpha=0.05$ the best method has a power of about $80\%$. In the goodness-of-fit problem the sample size differs between case studies, the values are in the data set \emph{case\textunderscore studies\textunderscore sample\textunderscore sizes} of \emph{Rgof}. In the two-sample problem both samples have size $500$. In half of the case studies the best method has a power between $73\%$ and $87\%$ (median about $80\%$), a few have a much lower power; in the case study betamixture1.est even the best method has a power of only $8\%$, so this case study contributes little to the comparison.

All results in this paper are for one configuration of each case study: one sample size, $\alpha=0.05$ and one parameter of the alternative. To check whether the rankings of the methods carry over to other configurations we used the 100 continuous two-sample case studies of \emph{R2sample}. For each case study, each level $\alpha=0.01, 0.05, 0.1$ and each target power of $50\%$, $70\%$ and $90\%$ we found the sample size at which the best method has the target power (median sample sizes between 186 and 829), and then estimated the powers of all methods at that sample size twice, with different random numbers. At a fixed power the three levels lead to different sample sizes, so this separates the power level from the sample size and the level. The results are in tables \ref{tblRank} and \ref{tblRankMeans}. Changing $\alpha$ from 0.01 to 0.1 at a fixed power (and so roughly halving the sample size), or the power from $50\%$ to $90\%$ at a fixed level, gives a mean Kendall rank correlation between the powers of the methods of 0.80 to 0.86, compared to 0.87 to 0.92 between two replications of the same configuration. The best method of one configuration is within three percentage points of the best method of the other in $89\%$ to $97\%$ of the case studies (replications: $95\%$ to $98\%$). In all nine configurations ZC and ZA have the highest mean power, followed closely by Epps-Singleton and then by HHG and ZK. Less stable is the set of methods within $90\%$ of the best (overlap $65\%$ to $85\%$, replications $82\%$ to $89\%$), as is to be expected of a set defined by a threshold. For alternatives that the best method detects with a power between $50\%$ and $90\%$ the rankings therefore hardly depend on the sample size, the level or the power. When all methods have a low power the rankings are dominated by Monte Carlo error.

\begin{table}[H]
\centering
\begingroup\small
\begin{tabular}{p{5.2cm}lrrrr}
  \hline
Configurations compared & Median n & Cases & Kendall tau & Best within 3 points & Overlap within 90\% \\ 
  \hline
power 50\%: alpha 0.01 vs 0.1 & 409 vs 186 & 100 & 0.80 & 0.89 & 0.70 \\ 
  power 70\%: alpha 0.01 vs 0.1 & 570 vs 296 & 100 & 0.86 & 0.97 & 0.80 \\ 
  power 90\%: alpha 0.01 vs 0.1 & 829 vs 482 & 100 & 0.86 & 0.97 & 0.85 \\ 
  alpha 0.01: power 50\% vs 90\% & 409 vs 829 & 100 & 0.86 & 0.90 & 0.65 \\ 
  alpha 0.05: power 50\% vs 90\% & 258 vs 600 & 100 & 0.85 & 0.93 & 0.73 \\ 
  alpha 0.1: power 50\% vs 90\% & 186 vs 482 & 100 & 0.83 & 0.97 & 0.71 \\ 
  replications, power 50\% (mean over alpha) & 409, 258, 186 & 100 & 0.87 & 0.95 & 0.82 \\ 
  replications, power 70\% (mean over alpha) & 570, 382, 296 & 100 & 0.90 & 0.96 & 0.87 \\ 
  replications, power 90\% (mean over alpha) & 829, 600, 482 & 100 & 0.92 & 0.98 & 0.89 \\ 
   \hline
\end{tabular}
\endgroup
\caption{Comparison of the rankings of the methods for the
          continuous two-sample case studies of R2sample in configurations with different level and target power (power of the best method). For each configuration the sample size was chosen so that the best method has the target power. Cases: number of case studies for which the target power was reached in both configurations (for the replications the mean over the three levels). Kendall tau: mean Kendall rank correlation of the powers of the methods; Best within 3 points: proportion of case studies in which the best method of the first configuration is within 3 percentage points of the best method of the second; Overlap within 90\%: mean overlap (Jaccard index) of the sets of methods with a power within 90\% of the best. Replications: two runs of the same configuration with different random numbers.} 
\label{tblRank}
\end{table}

\begin{table}[H]
\centering
\begingroup\small
\begin{tabular}{rrrrrrrrr}
  \hline
 & ZC & ZA & EppsS & HHG & ZK & EP small & ES small & Kuiper \\ 
  \hline
$\alpha=0.01$, 50\% & 35.6 & 36.1 & 35.2 & 28.6 & 27.4 & 19.4 & 19.4 & 18.7 \\ 
  $\alpha=0.01$, 70\% & 55.2 & 55.4 & 52.4 & 46.0 & 43.3 & 32.0 & 32.1 & 29.7 \\ 
  $\alpha=0.01$, 90\% & 76.6 & 76.5 & 72.3 & 67.7 & 63.6 & 50.5 & 49.8 & 45.6 \\ 
  $\alpha=0.05$, 50\% & 40.1 & 40.6 & 38.9 & 32.5 & 33.8 & 24.4 & 24.5 & 24.5 \\ 
  $\alpha=0.05$, 70\% & 57.8 & 58.0 & 54.7 & 48.1 & 48.5 & 35.9 & 35.8 & 35.3 \\ 
  $\alpha=0.05$, 90\% & 80.2 & 79.9 & 74.6 & 71.1 & 70.4 & 55.4 & 54.9 & 52.8 \\ 
  $\alpha=0.1$, 50\% & 42.1 & 42.4 & 40.8 & 35.1 & 36.9 & 28.0 & 28.2 & 27.8 \\ 
  $\alpha=0.1$, 70\% & 60.2 & 60.4 & 56.5 & 51.1 & 52.9 & 40.3 & 40.3 & 39.6 \\ 
  $\alpha=0.1$, 90\% & 81.7 & 81.3 & 76.1 & 72.2 & 73.8 & 58.3 & 57.8 & 56.4 \\ 
   \hline
\end{tabular}
\endgroup
\caption{Mean power (in percent, average of the two replications) of the eight methods with the highest mean power at $\alpha=0.05$ and a target power of 90\%, 100 continuous two-sample case studies, by level and target power.} 
\label{tblRankMeans}
\end{table}

\subsection{Computational details}\label{sec:compdetails}

All power studies use $B=1000$ simulation runs and $\alpha=0.05$; so the Monte Carlo standard error of a power estimate $p$ is $\sqrt{p(1-p)/1000}$, at most $1.6\%$. Each case study was run once with the parameter of the alternative and once with the parameter of the null hypothesis, which gives the true type I error of every method in every case study, see section \ref{sec:typeI}.

The results were computed with \emph{Rgof} 4.0.1 and \emph{R2sample} 5.1.0 under R 4.6.0. Every power study (one case study, type and data type, with the alternative or with the null hypothesis) uses its own random seed, so any single result can be reproduced with the routine \emph{gof\textunderscore power} or \emph{twosample\textunderscore power} of the packages; the seeds and the rule by which they were assigned, the settings and the software versions are stored with the results (attr(power\textunderscore study, "info")). The script that produces every table and figure of this paper, \emph{replication\textunderscore material\textunderscore 1D.Rmd}, is part of the replication material of this paper.

Table \ref{tblTime} gives the time for one complete test of all methods on one core, with p values found from $B=1000$ simulated or permuted data sets. In the power studies no nested simulation is needed: the test statistics of the $B=1000$ data sets simulated under the alternative are compared with critical values found from $B=1000$ data sets simulated under the null hypothesis.

\begin{table}[H]
\centering
\begingroup\small
\begin{tabular}{p{5cm}p{4.5cm}r}
  \hline
Problem & Size & Seconds \\ 
  \hline
Goodness-of-fit, continuous & n=300 & 0.12 \\ 
  Goodness-of-fit, continuous, estimation & n=200 & 0.13 \\ 
  Goodness-of-fit, discrete & n=300, 50 values & 0.033 \\ 
  Hybrid, nMC=1, continuous & n=300 & 0.51 \\ 
  Hybrid, nMC=5, continuous & n=300 & 1.1 \\ 
  Hybrid, nMC=1, discrete & n=300, 50 values & 1 \\ 
  Hybrid, nMC=5, discrete & n=300, 50 values & 1.1 \\ 
  Two-sample, continuous & n=m=500 & 0.68 \\ 
  Two-sample, discrete & n=500, m=500, 50 values & 1.1 \\ 
   \hline
\end{tabular}
\endgroup
\caption{Computing time in seconds for one complete test of all methods on one core, with p values found from B=1000 simulated or permuted data sets.} 
\label{tblTime}
\end{table}

All computations were done on a PC with an Intel Core i9-14900F processor (24 cores) and 32 GB of memory, running Windows 11 and R 4.6.0. The times in table \ref{tblTime} are for a single core. The power studies were run in parallel on 23 cores.

\subsection{Naming conventions}

A typical case study of type mixed in the goodness-of-fit problem is called dist1.dist2, where dist1 is the distribution under the null hypothesis and dist2 is the true distribution of the data, for example exponential.weibull. If parameters are estimated the name ends in .est. The case studies of the types mean to kurtosis are named after the distribution under the null hypothesis, for example normal-mix2. In the summary tables the type is added to the name, as in normal-mix2:variance.

\subsection{Examples}

\subsubsection{Example 1}

\begin{figure}[H]
\centering
\includegraphics[width=0.8\textwidth]{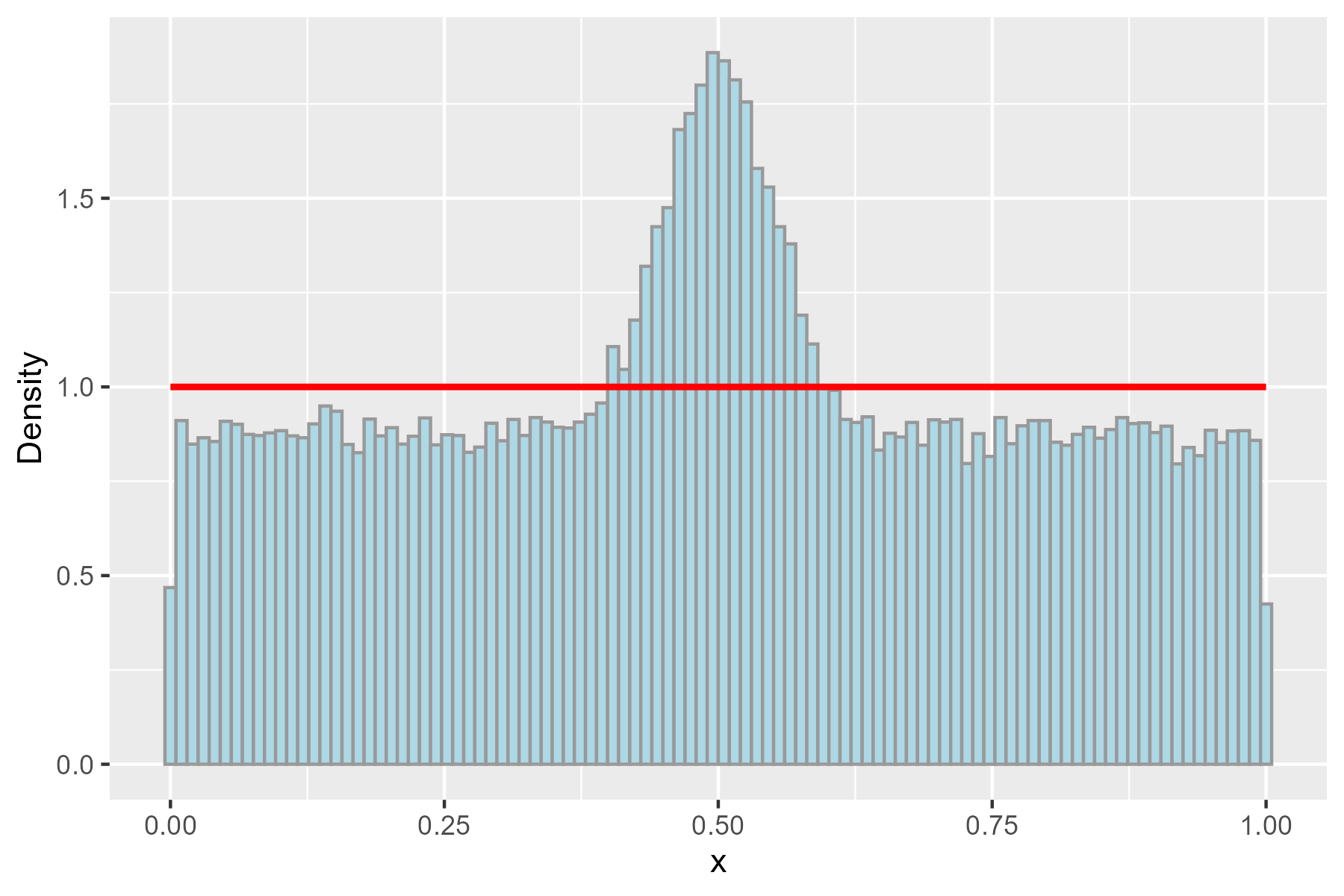}
\caption{Example 1: histogram of a large data set under the alternative and the density of the null hypothesis.}
\label{fig:ex1}
\end{figure}

The case study uniform-bump of type mixed for the goodness-of-fit problem with continuous data has under the null hypothesis a uniform distribution on $[0,1]$. Under the alternative a fraction of the observations come from a normal distribution with mean $0.5$ and a small standard deviation, so there is a narrow bump in the middle. Figure \ref{fig:ex1} shows a histogram of a large data set under the alternative, together with the density under the null hypothesis.

The results for this case study are recorded in the data set \emph{power\textunderscore study} included in the $\mathbf{R}$ package \emph{Rgof}. It is a list with elements \emph{cont} and \emph{disc}, each a list of matrices of powers, one for each type of case study, with the case studies in the rows and the methods in the columns. The true type I errors are in attr(power\textunderscore study, "Null"). The powers for this case study are in table \ref{tblEx1}.

\begin{table}[H]
\centering
\begin{tabular}{lllllllll}
  \hline
  \hline
W & K & ES-s-P & EP-s-P & ES-s-L & EP-s-L & ES-l-P & EP-l-P & KS \\ 
  83 & 81 & 73 & 73 & 70 & 70 & 47 & 47 & 44 \\ 
  ----- & ----- & ----- & ----- & ----- & ----- & ----- & ----- & ----- \\ 
  ES-l-L & EP-l-L & CvM & AD & ZK & Wassp1 & ZA & ZC &  \\ 
  35 & 35 & 30 & 28 & 26 & 22 & 21 & 18 &  \\ 
   \hline
\end{tabular}
\caption{Powers of various methods for case study uniform-bump (goodness-of-fit problem), continuous data, no parameter estimation} 
\label{tblEx1}
\end{table}

So Watson's test and Kuiper's test are best with a power of about $82\%$, followed by the chi-square tests with a small number of bins. The Zhang tests, overall the best methods, have a power of about $20\%$ here. A narrow bump in the middle of the distribution hardly changes the cdf in its tails, where these tests put their weight.

\subsubsection{Example 2}

\begin{figure}[H]
\centering
\includegraphics[width=0.8\textwidth]{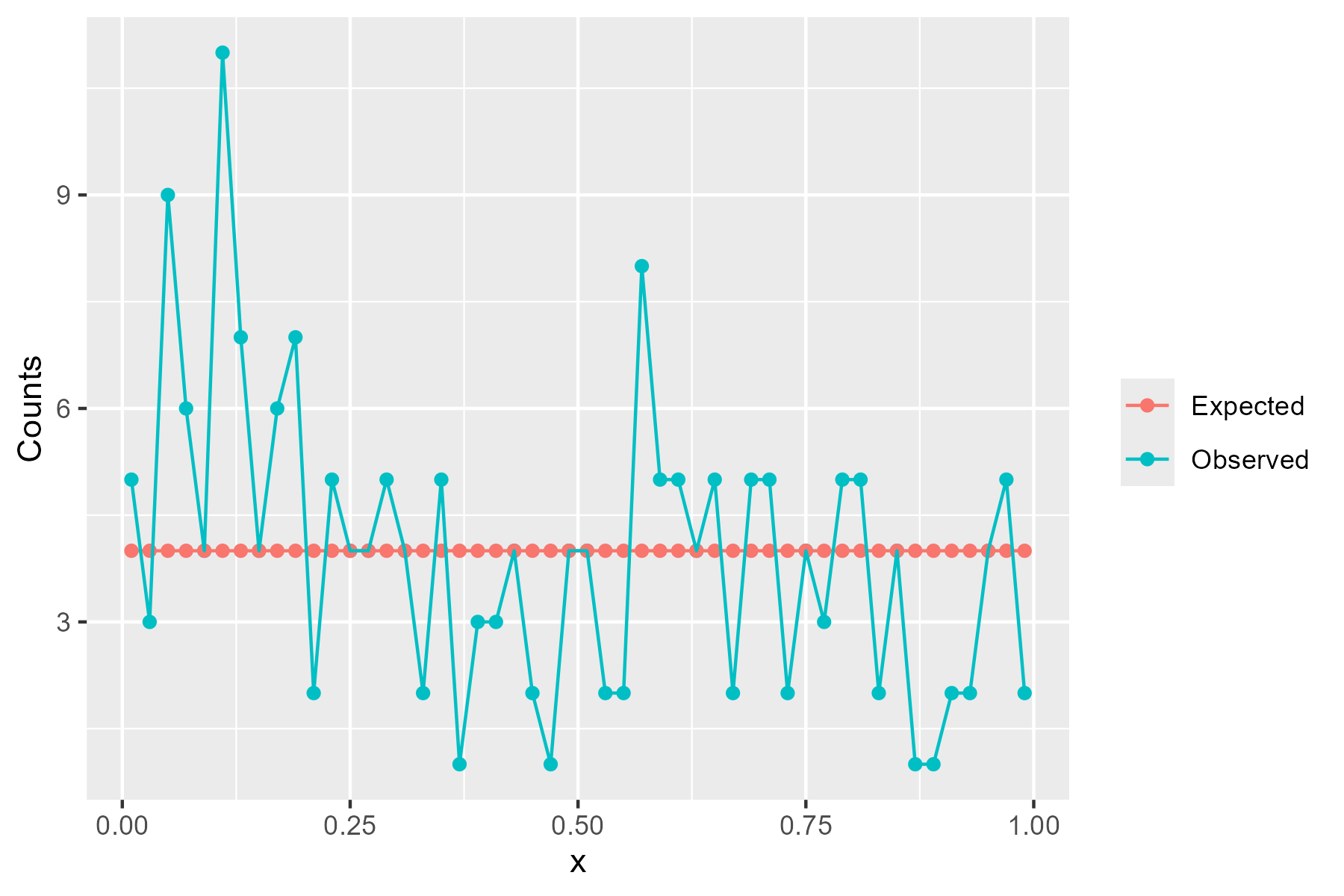}
\caption{Example 2: observed counts of a data set under the alternative and expected counts under the null hypothesis.}
\label{fig:ex2}
\end{figure}

As an example for discrete data we use the case study uniform-sine of type mixed. As with all the case studies for discrete data, the data of the corresponding continuous case study is binned into 50 bins. Under the alternative the density is a uniform density with an added sine wave. Figure \ref{fig:ex2} shows the observed counts of a data set under the alternative and the expected counts under the null hypothesis.

Table \ref{tblEx2} has the powers. Here the smooth test NS-A is best ($80\%$), followed by the chi-square tests ($72\%$ and $62\%$). The oscillation of the density changes the cdf very little, and so all the methods based on the edf have low power.

\begin{table}[H]
\centering
\begin{tabular}{lllllllll}
  \hline
  \hline
NS-A & s-P & l-P & BJ & NS-B & AD & KS & K & W \\ 
  80 & 72 & 62 & 48 & 43 & 42 & 42 & 39 & 38 \\ 
  ----- & ----- & ----- & ----- & ----- & ----- & ----- & ----- & ----- \\ 
  CvM & Wassp1 &  &  &  &  &  &  &  \\ 
  35 & 33 &  &  &  &  &  &  &  \\ 
   \hline
\end{tabular}
\caption{Powers of various methods for case study uniform-sine (goodness-of-fit problem), discrete data, no parameter estimation} 
\label{tblEx2}
\end{table}

\subsubsection{Example 3}

\begin{figure}[H]
\centering
\includegraphics[width=0.8\textwidth]{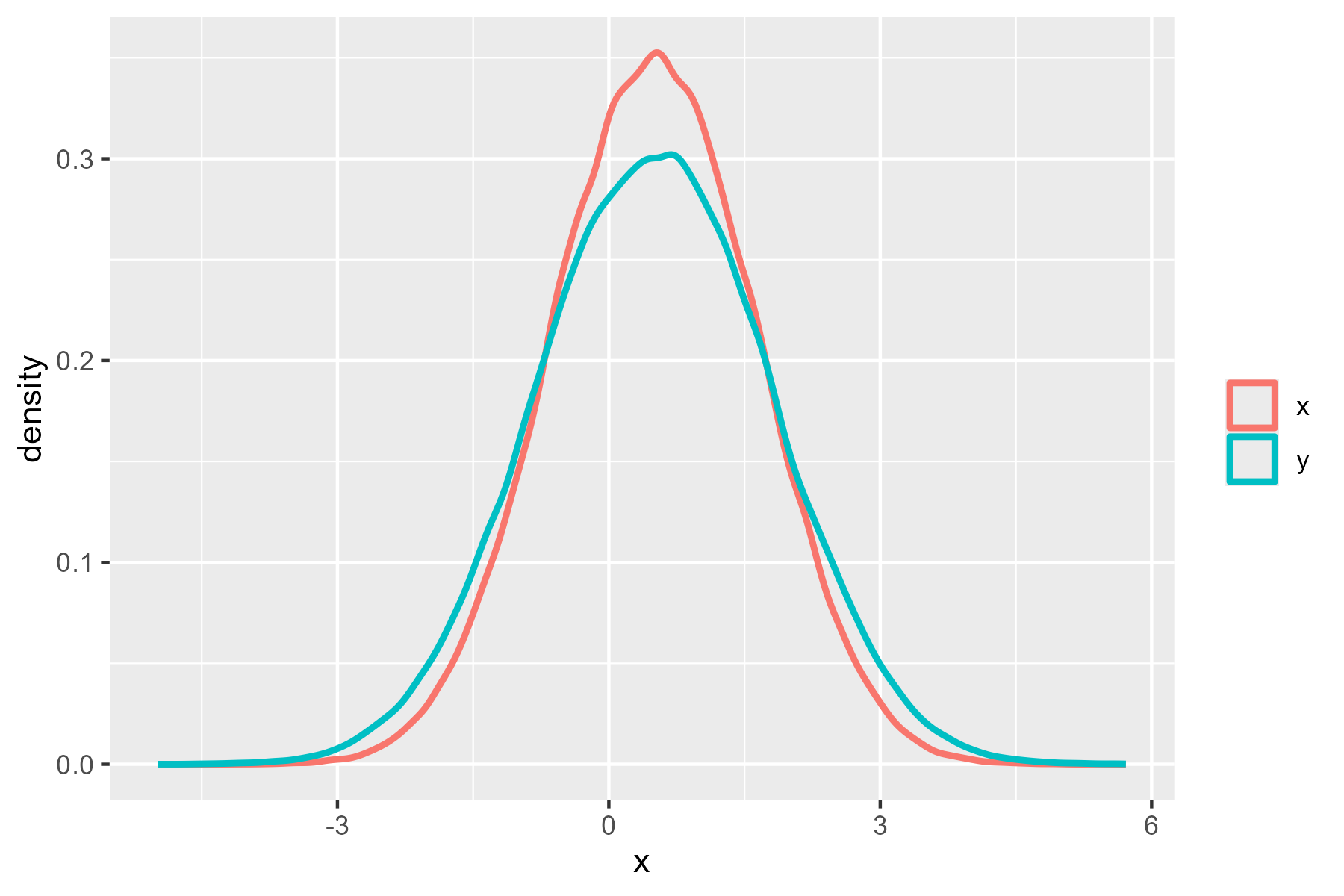}
\caption{Example 3: density estimates of large data sets from the two distributions.}
\label{fig:ex3}
\end{figure}

As an example of a two-sample problem we use the case study normal-mix2 of type variance. Here both distributions are mixtures of normal distributions with the same mean, skewness and kurtosis, they differ only in the variance. Figure \ref{fig:ex3} shows density estimates of two large data sets.

The results of the power study are in table \ref{tblEx3}. Here the Epps-Singleton test and the Zhang tests ZC and ZA are best, with a power of about $75\%$, followed by the chi-square test with a small number of equal-size bins ($59\%$). The Kolmogorov-Smirnov and Cram\'er-von Mises tests, which are most sensitive to differences in the center of the distribution, have a power below $20\%$.

\begin{table}[H]
\centering
\begin{tabular}{lllllllll}
  \hline
  \hline
EppsS & ZC & ZA & ES small & ZK & Kuiper & HHG & Wassp1 & EP small \\ 
  77 & 75 & 74 & 59 & 57 & 55 & 55 & 47 & 47 \\ 
  ----- & ----- & ----- & ----- & ----- & ----- & ----- & ----- & ----- \\ 
  AD & ES large & EP large & CvM & KS &  &  &  &  \\ 
  39 & 33 & 27 & 16 & 14 &  &  &  &  \\ 
   \hline
\end{tabular}
\caption{Powers of various methods for case study normal-mix2 with differing variances (two-sample problem), continuous data} 
\label{tblEx3}
\end{table}

The results of the power studies for the two-sample problem are recorded in the data set \emph{power\textunderscore study}, part of \emph{R2sample}. It is a list with elements \emph{cont} and \emph{disc}, each a list of matrices of powers, one for each type of case study, as well as the matrix \emph{Null} with the true type I errors.

\newpage

\section{Results}

We will now give the results of the various studies. For each we present the following four tables:

\begin{enumerate}
\item
Mean power: the mean power of each method over the included case studies.
\item
Percentage close to best: the percentage of case studies in which a method had a power within $90\%$ of the best method.
\item
Mean shortfall from best: the mean, over the case studies, of the power of the best method minus the power of the method, in percentage points.
\item
Best selection(s): for the combination of all case studies we find the smallest selection of methods such that for each study at least one of the methods has a power within $90\%$ of the best. Usually there is more than one such selection.
\end{enumerate}

Because in each case study the alternative was chosen so that the best method has a power of about $80\%$, the mean powers depend on this choice: they are best read relative to the best method of each case study, not as absolute measures of power. The mean shortfall and the performance profiles compare each method directly with the best one. The best selections are found by an exhaustive search over all subsets of the methods, starting with the smallest; all selections of the smallest size are reported, in increasing order of the mean power of their methods, so the last one listed has the highest mean power; this is the one used in the cross-validation of section \ref{sec:loo}. We did not examine thresholds other than $90\%$ for the selections.

The first two of these depend on the arbitrary choice of $90\%$. For the main studies we therefore also show performance profiles \cite{dolan2002}: for each method the share of case studies in which its power is at least $\tau$ times the power of the best method, as a function of $\tau$. A method whose curve is higher is closer to the best in more case studies; the value at $\tau=0.9$ is the percentage close to best of the tables, the value at $\tau=1$ the share of case studies in which the method is best.

The best selection rewards a method that is best in a single case study, even if it is weak otherwise. It should therefore be read together with the mean powers and the shortfalls, and in section \ref{sec:loo} we check by cross-validation how well the selections carry over to case studies that were not used to find them.

In a few case studies some methods could not be calculated (for example the Anderson-Darling test for the noncentral chi-square distribution). In the summaries these methods are given a power of 0 in those case studies.

\subsection{Goodness-of-fit}

\subsubsection{Continuous data, all case studies}

\begin{table}[H]
\centering
\caption{Goodness-of-fit, continuous data, all case studies}
\label{tab:GC}

\begin{tabular}{lllllllll}
\toprule
\multicolumn{9}{l}{\textbf{Mean Power}} \\
\midrule
  \hline
  \hline
ZC & ZA & ZK & AD & W & K & Wassp1 & ES-s-L & EP-s-L \\ 
  69 & 66 & 62 & 51 & 50 & 49 & 45 & 44 & 43 \\ 
  ----- & ----- & ----- & ----- & ----- & ----- & ----- & ----- & ----- \\ 
  ES-s-P & EP-s-P & CvM & KS & ES-l-P & ES-l-L & EP-l-P & EP-l-L &  \\ 
  43 & 43 & 42 & 40 & 35 & 35 & 33 & 33 &  \\ 
   \hline
\addlinespace
\multicolumn{9}{l}{\textbf{Percentage close to best}} \\
\midrule
  \hline
  \hline
ZC & ZA & AD & Wassp1 & W & CvM & K & ZK & ES-s-L \\ 
  61 & 49 & 29 & 26 & 25 & 21 & 19 & 16 & 12 \\ 
  ----- & ----- & ----- & ----- & ----- & ----- & ----- & ----- & ----- \\ 
  ES-s-P & EP-s-L & EP-s-P & KS & ES-l-P & EP-l-P & ES-l-L & EP-l-L &  \\ 
  11 & 9 & 8 & 8 & 2 & 2 & 2 & 2 &  \\ 
   \hline
\addlinespace
\multicolumn{9}{l}{\textbf{Mean shortfall from best}} \\
\midrule
  \hline
  \hline
ZC & ZA & ZK & AD & W & K & Wassp1 & ES-s-L & EP-s-L \\ 
  12 & 14 & 18 & 29 & 30 & 31 & 35 & 36 & 37 \\ 
  ----- & ----- & ----- & ----- & ----- & ----- & ----- & ----- & ----- \\ 
  ES-s-P & EP-s-P & CvM & KS & ES-l-P & ES-l-L & EP-l-P & EP-l-L &  \\ 
  37 & 38 & 38 & 40 & 45 & 45 & 47 & 47 &  \\ 
   \hline
\addlinespace
\multicolumn{9}{l}{\textbf{Best selection(s)}} \\
\midrule
  \hline
  \hline
CvM & EP-s-P & W & Wassp1 & ZA & ZC \\ 
  CvM & ES-s-P & W & Wassp1 & ZA & ZC \\ 
  CvM & EP-s-L & W & Wassp1 & ZA & ZC \\ 
  CvM & ES-s-L & W & Wassp1 & ZA & ZC \\ 
  AD & EP-s-P & W & Wassp1 & ZA & ZC \\ 
  AD & ES-s-P & W & Wassp1 & ZA & ZC \\ 
  AD & EP-s-L & W & Wassp1 & ZA & ZC \\ 
  AD & ES-s-L & W & Wassp1 & ZA & ZC \\ 
   \hline
\bottomrule
\end{tabular}
\end{table}

\begin{figure}[H]
\centering
\includegraphics[width=0.85\textwidth]{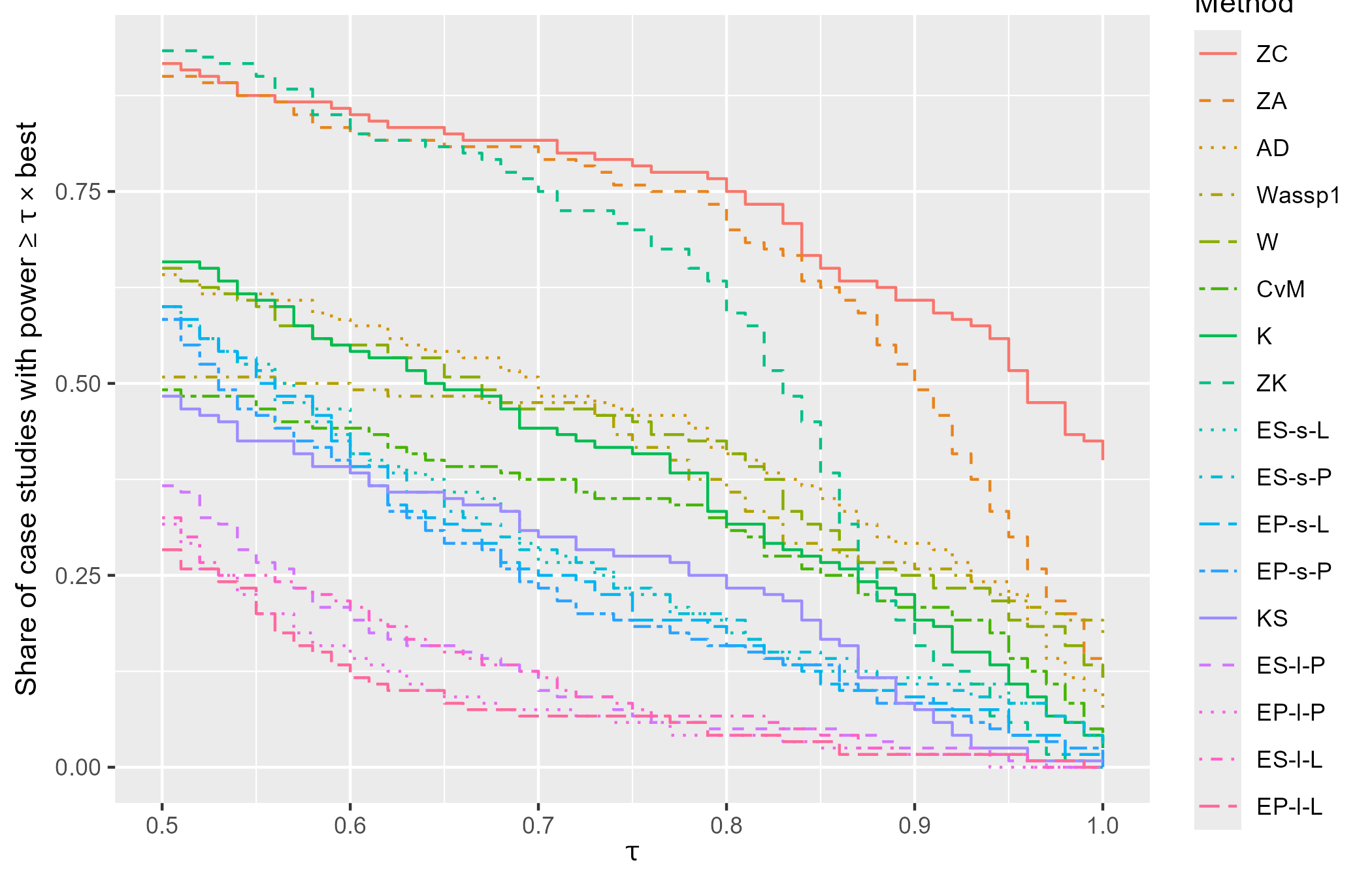}
\caption{Performance profiles, goodness-of-fit, continuous data.}
\label{fig:profGC}
\end{figure}

Zhang's tests ZC and ZA have by far the highest mean power, but six methods are needed to always have one close to the best: ZC, ZA, Watson, Wasserstein, Anderson-Darling or Cram\'er-von Mises and a chi-square test with a small number of bins.

\subsubsection{Continuous data, by type}

\begin{table}[H]
\centering
\caption{Goodness-of-fit, continuous data, case studies of type mean}
\label{tab:GCmean}

\begin{tabular}{lllllllll}
\toprule
\multicolumn{9}{l}{\textbf{Mean Power}} \\
\midrule
  \hline
  \hline
ZC & ZA & ZK & AD & CvM & Wassp1 & KS & EP-s-L & ES-s-L \\ 
  76 & 75 & 72 & 68 & 64 & 61 & 57 & 49 & 47 \\ 
  ----- & ----- & ----- & ----- & ----- & ----- & ----- & ----- & ----- \\ 
  EP-s-P & ES-s-P & K & W & ES-l-P & ES-l-L & EP-l-L & EP-l-P &  \\ 
  47 & 47 & 38 & 36 & 35 & 35 & 33 & 33 &  \\ 
   \hline
\addlinespace
\multicolumn{9}{l}{\textbf{Percentage close to best}} \\
\midrule
  \hline
  \hline
ZA & ZC & AD & CvM & ZK & Wassp1 & KS & W & K \\ 
  75 & 70 & 50 & 40 & 30 & 30 & 15 & 10 & 5 \\ 
  ----- & ----- & ----- & ----- & ----- & ----- & ----- & ----- & ----- \\ 
  ES-s-P & EP-s-P & ES-s-L & EP-s-L & ES-l-P & EP-l-P & ES-l-L & EP-l-L &  \\ 
  5 & 5 & 5 & 5 & 0 & 0 & 0 & 0 &  \\ 
   \hline
\addlinespace
\multicolumn{9}{l}{\textbf{Mean shortfall from best}} \\
\midrule
  \hline
  \hline
ZC & ZA & ZK & AD & CvM & Wassp1 & KS & EP-s-L & ES-s-L \\ 
  7 & 8 & 11 & 15 & 20 & 22 & 26 & 34 & 36 \\ 
  ----- & ----- & ----- & ----- & ----- & ----- & ----- & ----- & ----- \\ 
  EP-s-P & ES-s-P & K & W & ES-l-P & ES-l-L & EP-l-L & EP-l-P &  \\ 
  36 & 37 & 45 & 48 & 48 & 49 & 50 & 50 &  \\ 
   \hline
\addlinespace
\multicolumn{9}{l}{\textbf{Best selection(s)}} \\
\midrule
  \hline
  \hline
W & Wassp1 & ZA \\ 
  KS & Wassp1 & ZA \\ 
  CvM & Wassp1 & ZA \\ 
   \hline
\bottomrule
\end{tabular}
\end{table}

\begin{table}[H]
\centering
\caption{Goodness-of-fit, continuous data, case studies of type variance}
\label{tab:GCvariance}

\begin{tabular}{lllllllll}
\toprule
\multicolumn{9}{l}{\textbf{Mean Power}} \\
\midrule
  \hline
  \hline
ZC & ZA & ZK & AD & W & K & Wassp1 & ES-s-P & ES-s-L \\ 
  77 & 70 & 67 & 44 & 42 & 39 & 37 & 34 & 31 \\ 
  ----- & ----- & ----- & ----- & ----- & ----- & ----- & ----- & ----- \\ 
  ES-l-P & EP-l-P & EP-s-P & EP-s-L & ES-l-L & EP-l-L & CvM & KS &  \\ 
  30 & 30 & 29 & 29 & 28 & 28 & 26 & 23 &  \\ 
   \hline
\addlinespace
\multicolumn{9}{l}{\textbf{Percentage close to best}} \\
\midrule
  \hline
  \hline
ZC & ZA & ZK & Wassp1 & W & K & AD & CvM & KS \\ 
  90 & 45 & 25 & 20 & 15 & 5 & 5 & 5 & 0 \\ 
  ----- & ----- & ----- & ----- & ----- & ----- & ----- & ----- & ----- \\ 
  ES-l-P & ES-s-P & EP-l-P & EP-s-P & ES-l-L & ES-s-L & EP-l-L & EP-s-L &  \\ 
  0 & 0 & 0 & 0 & 0 & 0 & 0 & 0 &  \\ 
   \hline
\addlinespace
\multicolumn{9}{l}{\textbf{Mean shortfall from best}} \\
\midrule
  \hline
  \hline
ZC & ZA & ZK & AD & W & K & Wassp1 & ES-s-P & ES-s-L \\ 
  2 & 9 & 12 & 35 & 37 & 40 & 42 & 45 & 48 \\ 
  ----- & ----- & ----- & ----- & ----- & ----- & ----- & ----- & ----- \\ 
  ES-l-P & EP-l-P & EP-s-P & EP-s-L & ES-l-L & EP-l-L & CvM & KS &  \\ 
  49 & 49 & 50 & 50 & 51 & 51 & 53 & 56 &  \\ 
   \hline
\addlinespace
\multicolumn{9}{l}{\textbf{Best selection(s)}} \\
\midrule
  \hline
  \hline
CvM & Wassp1 & ZC \\ 
  W & Wassp1 & ZC \\ 
  AD & Wassp1 & ZC \\ 
   \hline
\bottomrule
\end{tabular}
\end{table}

\begin{table}[H]
\centering
\caption{Goodness-of-fit, continuous data, case studies of type skewness}
\label{tab:GCskewness}

\begin{tabular}{lllllllll}
\toprule
\multicolumn{9}{l}{\textbf{Mean Power}} \\
\midrule
  \hline
  \hline
ZA & W & ZC & K & EP-s-L & EP-s-P & ES-s-L & ES-s-P & ZK \\ 
  67 & 63 & 62 & 62 & 60 & 59 & 59 & 58 & 56 \\ 
  ----- & ----- & ----- & ----- & ----- & ----- & ----- & ----- & ----- \\ 
  ES-l-L & ES-l-P & EP-l-P & EP-l-L & AD & KS & Wassp1 & CvM &  \\ 
  49 & 47 & 43 & 42 & 42 & 34 & 33 & 32 &  \\ 
   \hline
\addlinespace
\multicolumn{9}{l}{\textbf{Percentage close to best}} \\
\midrule
  \hline
  \hline
ZA & ZC & ES-s-P & ES-s-L & W & EP-s-L & K & EP-s-P & AD \\ 
  65 & 50 & 35 & 35 & 30 & 25 & 20 & 20 & 10 \\ 
  ----- & ----- & ----- & ----- & ----- & ----- & ----- & ----- & ----- \\ 
  Wassp1 & CvM & ZK & KS & ES-l-P & EP-l-P & ES-l-L & EP-l-L &  \\ 
  10 & 5 & 5 & 0 & 0 & 0 & 0 & 0 &  \\ 
   \hline
\addlinespace
\multicolumn{9}{l}{\textbf{Mean shortfall from best}} \\
\midrule
  \hline
  \hline
ZA & W & ZC & K & EP-s-L & EP-s-P & ES-s-L & ES-s-P & ZK \\ 
  11 & 15 & 16 & 16 & 18 & 19 & 19 & 20 & 22 \\ 
  ----- & ----- & ----- & ----- & ----- & ----- & ----- & ----- & ----- \\ 
  ES-l-L & ES-l-P & EP-l-P & EP-l-L & AD & KS & Wassp1 & CvM &  \\ 
  30 & 31 & 35 & 36 & 37 & 44 & 45 & 46 &  \\ 
   \hline
\addlinespace
\multicolumn{9}{l}{\textbf{Best selection(s)}} \\
\midrule
  \hline
  \hline
CvM & ES-s-P & K & Wassp1 & ZA \\ 
  CvM & ES-s-L & K & Wassp1 & ZA \\ 
  CvM & EP-s-P & K & Wassp1 & ZA \\ 
  CvM & ES-s-P & W & Wassp1 & ZA \\ 
  CvM & EP-s-L & K & Wassp1 & ZA \\ 
  CvM & ES-s-L & W & Wassp1 & ZA \\ 
  CvM & EP-s-P & W & Wassp1 & ZA \\ 
  CvM & EP-s-L & W & Wassp1 & ZA \\ 
  AD & ES-s-P & K & Wassp1 & ZA \\ 
  AD & ES-s-L & K & Wassp1 & ZA \\ 
  AD & EP-s-P & K & Wassp1 & ZA \\ 
  AD & ES-s-P & W & Wassp1 & ZA \\ 
  AD & EP-s-L & K & Wassp1 & ZA \\ 
  AD & ES-s-L & W & Wassp1 & ZA \\ 
  AD & EP-s-P & W & Wassp1 & ZA \\ 
  AD & EP-s-L & W & Wassp1 & ZA \\ 
   \hline
\bottomrule
\end{tabular}
\end{table}

\begin{table}[H]
\centering
\caption{Goodness-of-fit, continuous data, case studies of type kurtosis}
\label{tab:GCkurtosis}

\begin{tabular}{lllllllll}
\toprule
\multicolumn{9}{l}{\textbf{Mean Power}} \\
\midrule
  \hline
  \hline
ZC & ZA & ZK & K & W & AD & ES-l-P & ES-l-L & EP-l-L \\ 
  74 & 68 & 66 & 42 & 41 & 37 & 36 & 36 & 35 \\ 
  ----- & ----- & ----- & ----- & ----- & ----- & ----- & ----- & ----- \\ 
  EP-l-P & ES-s-L & EP-s-L & EP-s-P & ES-s-P & Wassp1 & KS & CvM &  \\ 
  35 & 34 & 33 & 32 & 32 & 28 & 25 & 25 &  \\ 
   \hline
\addlinespace
\multicolumn{9}{l}{\textbf{Percentage close to best}} \\
\midrule
  \hline
  \hline
ZC & ZA & K & W & ZK & EP-s-P & EP-s-L & AD & CvM \\ 
  75 & 50 & 20 & 20 & 15 & 10 & 10 & 5 & 5 \\ 
  ----- & ----- & ----- & ----- & ----- & ----- & ----- & ----- & ----- \\ 
  Wassp1 & ES-l-P & ES-s-P & ES-l-L & ES-s-L & KS & EP-l-P & EP-l-L &  \\ 
  5 & 5 & 5 & 5 & 5 & 0 & 0 & 0 &  \\ 
   \hline
\addlinespace
\multicolumn{9}{l}{\textbf{Mean shortfall from best}} \\
\midrule
  \hline
  \hline
ZC & ZA & ZK & K & W & AD & ES-l-P & ES-l-L & EP-l-L \\ 
  8 & 15 & 16 & 40 & 41 & 45 & 46 & 47 & 47 \\ 
  ----- & ----- & ----- & ----- & ----- & ----- & ----- & ----- & ----- \\ 
  EP-l-P & ES-s-L & EP-s-L & EP-s-P & ES-s-P & Wassp1 & KS & CvM &  \\ 
  47 & 48 & 49 & 50 & 50 & 55 & 57 & 58 &  \\ 
   \hline
\addlinespace
\multicolumn{9}{l}{\textbf{Best selection(s)}} \\
\midrule
  \hline
  \hline
CvM & W & Wassp1 & ZC \\ 
  CvM & K & Wassp1 & ZC \\ 
  EP-s-P & W & Wassp1 & ZC \\ 
  EP-s-P & K & Wassp1 & ZC \\ 
  EP-s-L & W & Wassp1 & ZC \\ 
  EP-s-L & K & Wassp1 & ZC \\ 
  AD & W & Wassp1 & ZC \\ 
  AD & K & Wassp1 & ZC \\ 
   \hline
\bottomrule
\end{tabular}
\end{table}

\begin{table}[H]
\centering
\caption{Goodness-of-fit, continuous data, case studies of type mixed}
\label{tab:GCmixed}

\begin{tabular}{lllllllll}
\toprule
\multicolumn{9}{l}{\textbf{Mean Power}} \\
\midrule
  \hline
  \hline
ZC & W & ZA & AD & K & Wassp1 & ZK & CvM & KS \\ 
  60 & 60 & 59 & 59 & 57 & 57 & 56 & 53 & 50 \\ 
  ----- & ----- & ----- & ----- & ----- & ----- & ----- & ----- & ----- \\ 
  ES-s-L & ES-s-P & EP-s-L & EP-s-P & ES-l-L & ES-l-P & EP-l-L & EP-l-P &  \\ 
  46 & 45 & 45 & 44 & 32 & 32 & 30 & 29 &  \\ 
   \hline
\addlinespace
\multicolumn{9}{l}{\textbf{Percentage close to best}} \\
\midrule
  \hline
  \hline
AD & Wassp1 & ZC & W & CvM & K & ZA & KS & ES-s-L \\ 
  52 & 45 & 40 & 38 & 35 & 32 & 30 & 15 & 12 \\ 
  ----- & ----- & ----- & ----- & ----- & ----- & ----- & ----- & ----- \\ 
  ZK & ES-s-P & EP-s-P & EP-s-L & ES-l-P & EP-l-P & EP-l-L & ES-l-L &  \\ 
  10 & 10 & 8 & 8 & 5 & 5 & 5 & 2 &  \\ 
   \hline
\addlinespace
\multicolumn{9}{l}{\textbf{Mean shortfall from best}} \\
\midrule
  \hline
  \hline
ZC & W & ZA & AD & K & Wassp1 & ZK & CvM & KS \\ 
  19 & 19 & 20 & 20 & 22 & 22 & 23 & 26 & 29 \\ 
  ----- & ----- & ----- & ----- & ----- & ----- & ----- & ----- & ----- \\ 
  ES-s-L & ES-s-P & EP-s-L & EP-s-P & ES-l-L & ES-l-P & EP-l-L & EP-l-P &  \\ 
  33 & 34 & 34 & 35 & 47 & 47 & 49 & 50 &  \\ 
   \hline
\addlinespace
\multicolumn{9}{l}{\textbf{Best selection(s)}} \\
\midrule
  \hline
  \hline
EP-l-P & W & Wassp1 & ZA & ZC \\ 
  EP-l-L & W & Wassp1 & ZA & ZC \\ 
  ES-l-P & W & Wassp1 & ZA & ZC \\ 
  ES-l-L & W & Wassp1 & ZA & ZC \\ 
  EP-s-P & W & Wassp1 & ZA & ZC \\ 
  EP-s-L & W & Wassp1 & ZA & ZC \\ 
  ES-s-P & W & Wassp1 & ZA & ZC \\ 
  ES-s-L & W & Wassp1 & ZA & ZC \\ 
   \hline
\bottomrule
\end{tabular}
\end{table}

The type of the difference matters a great deal (Tables \ref{tab:GCmean} to \ref{tab:GCmixed}). If the distributions differ only in the variance or in the kurtosis, ZC is close to the best in most case studies. If they differ in the skewness ZA is best, but Watson's test, Kuiper's test and the chi-square tests with a small number of bins have about the same mean power. In the case studies of type mixed the differences between the methods are much smaller, here Watson, ZC, ZA, Anderson-Darling, Kuiper and Wasserstein all have a mean power of about $60\%$.

\subsubsection{Continuous data, with and without parameter estimation}

\begin{table}[H]
\centering
\caption{Goodness-of-fit, continuous data, case studies of type mixed without parameter estimation}
\label{tab:GCnE}

\begin{tabular}{lllllllll}
\toprule
\multicolumn{9}{l}{\textbf{Mean Power}} \\
\midrule
  \hline
  \hline
ZC & ZA & ZK & W & K & Wassp1 & AD & ES-s-L & ES-s-P \\ 
  65 & 63 & 60 & 55 & 55 & 54 & 53 & 50 & 50 \\ 
  ----- & ----- & ----- & ----- & ----- & ----- & ----- & ----- & ----- \\ 
  EP-s-L & EP-s-P & KS & CvM & ES-l-P & EP-l-P & ES-l-L & EP-l-L &  \\ 
  50 & 49 & 45 & 45 & 37 & 36 & 36 & 35 &  \\ 
   \hline
\addlinespace
\multicolumn{9}{l}{\textbf{Percentage close to best}} \\
\midrule
  \hline
  \hline
ZC & Wassp1 & K & AD & W & ZA & CvM & ZK & ES-s-P \\ 
  40 & 40 & 35 & 35 & 35 & 30 & 20 & 15 & 15 \\ 
  ----- & ----- & ----- & ----- & ----- & ----- & ----- & ----- & ----- \\ 
  ES-s-L & KS & ES-l-P & EP-l-P & EP-s-P & EP-s-L & ES-l-L & EP-l-L &  \\ 
  15 & 10 & 10 & 10 & 10 & 10 & 5 & 5 &  \\ 
   \hline
\addlinespace
\multicolumn{9}{l}{\textbf{Mean shortfall from best}} \\
\midrule
  \hline
  \hline
ZC & ZA & ZK & W & K & Wassp1 & AD & ES-s-L & ES-s-P \\ 
  17 & 19 & 22 & 26 & 27 & 28 & 29 & 32 & 32 \\ 
  ----- & ----- & ----- & ----- & ----- & ----- & ----- & ----- & ----- \\ 
  EP-s-L & EP-s-P & KS & CvM & ES-l-P & EP-l-P & ES-l-L & EP-l-L &  \\ 
  32 & 33 & 37 & 37 & 45 & 46 & 46 & 47 &  \\ 
   \hline
\addlinespace
\multicolumn{9}{l}{\textbf{Best selection(s)}} \\
\midrule
  \hline
  \hline
EP-l-L & K & Wassp1 & ZA & ZC \\ 
  EP-l-L & W & Wassp1 & ZA & ZC \\ 
  ES-l-L & K & Wassp1 & ZA & ZC \\ 
  EP-l-P & K & Wassp1 & ZA & ZC \\ 
  ES-l-L & W & Wassp1 & ZA & ZC \\ 
  EP-l-P & W & Wassp1 & ZA & ZC \\ 
  ES-l-P & K & Wassp1 & ZA & ZC \\ 
  ES-l-P & W & Wassp1 & ZA & ZC \\ 
  EP-s-P & K & Wassp1 & ZA & ZC \\ 
  EP-s-L & K & Wassp1 & ZA & ZC \\ 
  ES-s-P & K & Wassp1 & ZA & ZC \\ 
  EP-s-P & W & Wassp1 & ZA & ZC \\ 
  EP-s-L & W & Wassp1 & ZA & ZC \\ 
  ES-s-L & K & Wassp1 & ZA & ZC \\ 
  ES-s-P & W & Wassp1 & ZA & ZC \\ 
  ES-s-L & W & Wassp1 & ZA & ZC \\ 
   \hline
\bottomrule
\end{tabular}
\end{table}

\begin{table}[H]
\centering
\caption{Goodness-of-fit, continuous data, case studies of type mixed with parameter estimation}
\label{tab:GCE}

\begin{tabular}{lllllllll}
\toprule
\multicolumn{9}{l}{\textbf{Mean Power}} \\
\midrule
  \hline
  \hline
W & AD & CvM & K & Wassp1 & ZC & KS & ZA & ZK \\ 
  65 & 64 & 61 & 60 & 59 & 56 & 56 & 55 & 52 \\ 
  ----- & ----- & ----- & ----- & ----- & ----- & ----- & ----- & ----- \\ 
  ES-s-L & EP-s-L & ES-s-P & EP-s-P & ES-l-L & ES-l-P & EP-l-L & EP-l-P &  \\ 
  41 & 40 & 40 & 39 & 29 & 26 & 26 & 22 &  \\ 
   \hline
\addlinespace
\multicolumn{9}{l}{\textbf{Percentage close to best}} \\
\midrule
  \hline
  \hline
AD & CvM & Wassp1 & W & ZC & K & ZA & KS & ES-s-L \\ 
  70 & 50 & 50 & 40 & 40 & 30 & 30 & 20 & 10 \\ 
  ----- & ----- & ----- & ----- & ----- & ----- & ----- & ----- & ----- \\ 
  ZK & ES-s-P & EP-s-P & EP-l-L & EP-s-L & ES-l-P & EP-l-P & ES-l-L &  \\ 
  5 & 5 & 5 & 5 & 5 & 0 & 0 & 0 &  \\ 
   \hline
\addlinespace
\multicolumn{9}{l}{\textbf{Mean shortfall from best}} \\
\midrule
  \hline
  \hline
W & AD & CvM & K & Wassp1 & ZC & KS & ZA & ZK \\ 
  12 & 12 & 15 & 16 & 17 & 20 & 20 & 21 & 25 \\ 
  ----- & ----- & ----- & ----- & ----- & ----- & ----- & ----- & ----- \\ 
  ES-s-L & EP-s-L & ES-s-P & EP-s-P & ES-l-L & ES-l-P & EP-l-L & EP-l-P &  \\ 
  35 & 36 & 36 & 37 & 48 & 50 & 50 & 54 &  \\ 
   \hline
\addlinespace
\multicolumn{9}{l}{\textbf{Best selection(s)}} \\
\midrule
  \hline
  \hline
W & Wassp1 & ZC \\ 
   \hline
\bottomrule
\end{tabular}
\end{table}

With parameter estimation the Zhang tests lose their advantage (Table \ref{tab:GCE}, without estimation Table \ref{tab:GCnE}). Watson's test and the Anderson-Darling test then have the highest mean power, and only three methods are needed, Watson, ZC and Wasserstein, compared to five for the case studies of type mixed without parameter estimation.

\subsubsection{Discrete data, all case studies}

\begin{table}[H]
\centering
\caption{Goodness-of-fit, discrete data, all case studies}
\label{tab:GD}

\begin{tabular}{llllllll}
\toprule
\multicolumn{8}{l}{\textbf{Mean Power}} \\
\midrule
  \hline
  \hline
NS-B & NS-A & BJ & K & AD & W & s-P & l-P \\ 
  71 & 68 & 66 & 60 & 59 & 59 & 56 & 52 \\ 
  ----- & ----- & ----- & ----- & ----- & ----- & ----- & ----- \\ 
  Wassp1 & CvM & KS &  &  &  &  &  \\ 
  46 & 43 & 43 &  &  &  &  &  \\ 
   \hline
\addlinespace
\multicolumn{8}{l}{\textbf{Percentage close to best}} \\
\midrule
  \hline
  \hline
NS-A & NS-B & BJ & W & AD & K & s-P & CvM \\ 
  55 & 43 & 33 & 32 & 31 & 23 & 21 & 20 \\ 
  ----- & ----- & ----- & ----- & ----- & ----- & ----- & ----- \\ 
  Wassp1 & KS & l-P &  &  &  &  &  \\ 
  20 & 8 & 4 &  &  &  &  &  \\ 
   \hline
\addlinespace
\multicolumn{8}{l}{\textbf{Mean shortfall from best}} \\
\midrule
  \hline
  \hline
NS-B & NS-A & BJ & K & AD & W & s-P & l-P \\ 
  12 & 15 & 17 & 23 & 24 & 24 & 27 & 31 \\ 
  ----- & ----- & ----- & ----- & ----- & ----- & ----- & ----- \\ 
  Wassp1 & CvM & KS &  &  &  &  &  \\ 
  38 & 40 & 41 &  &  &  &  &  \\ 
   \hline
\addlinespace
\multicolumn{8}{l}{\textbf{Best selection(s)}} \\
\midrule
  \hline
  \hline
AD & BJ & NS-A & NS-B & W & Wassp1 \\ 
  AD & BJ & K & NS-A & NS-B & Wassp1 \\ 
   \hline
\bottomrule
\end{tabular}
\end{table}

\begin{figure}[H]
\centering
\includegraphics[width=0.85\textwidth]{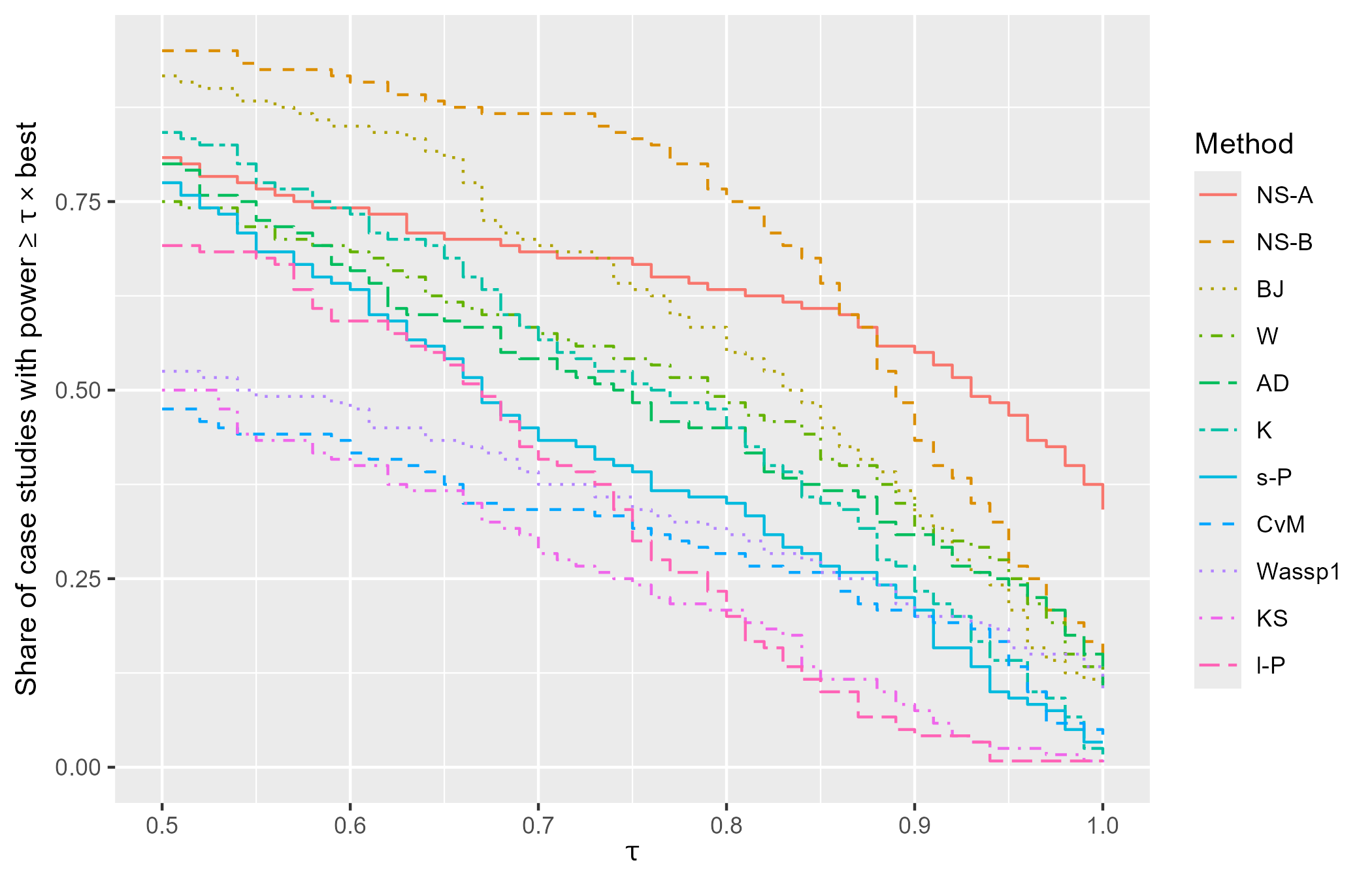}
\caption{Performance profiles, goodness-of-fit, discrete data.}
\label{fig:profGD}
\end{figure}

The smooth tests NS-B and NS-A and the Berk-Jones test have the highest mean power, followed by Kuiper, Anderson-Darling and Watson. The best selection consists of NS-B, NS-A, Berk-Jones, Anderson-Darling, Wasserstein and Watson (or Kuiper); the chi-square tests are no longer needed.

\subsubsection{Discrete data, by type}

\begin{table}[H]
\centering
\caption{Goodness-of-fit, discrete data, case studies of type mean}
\label{tab:GDmean}

\begin{tabular}{llllllll}
\toprule
\multicolumn{8}{l}{\textbf{Mean Power}} \\
\midrule
  \hline
  \hline
AD & NS-B & BJ & CvM & Wassp1 & KS & NS-A & s-P \\ 
  72 & 67 & 67 & 62 & 57 & 55 & 54 & 53 \\ 
  ----- & ----- & ----- & ----- & ----- & ----- & ----- & ----- \\ 
  l-P & K & W &  &  &  &  &  \\ 
  46 & 40 & 34 &  &  &  &  &  \\ 
   \hline
\addlinespace
\multicolumn{8}{l}{\textbf{Percentage close to best}} \\
\midrule
  \hline
  \hline
AD & BJ & NS-A & CvM & NS-B & Wassp1 & KS & s-P \\ 
  80 & 50 & 40 & 35 & 30 & 25 & 20 & 20 \\ 
  ----- & ----- & ----- & ----- & ----- & ----- & ----- & ----- \\ 
  W & K & l-P &  &  &  &  &  \\ 
  15 & 10 & 0 &  &  &  &  &  \\ 
   \hline
\addlinespace
\multicolumn{8}{l}{\textbf{Mean shortfall from best}} \\
\midrule
  \hline
  \hline
AD & NS-B & BJ & CvM & Wassp1 & KS & NS-A & s-P \\ 
  4 & 9 & 10 & 14 & 19 & 22 & 23 & 23 \\ 
  ----- & ----- & ----- & ----- & ----- & ----- & ----- & ----- \\ 
  l-P & K & W &  &  &  &  &  \\ 
  30 & 37 & 42 &  &  &  &  &  \\ 
   \hline
\addlinespace
\multicolumn{8}{l}{\textbf{Best selection(s)}} \\
\midrule
  \hline
  \hline
AD & NS-A & s-P \\ 
  AD & BJ & s-P \\ 
  AD & NS-A & NS-B \\ 
  AD & BJ & NS-B \\ 
   \hline
\bottomrule
\end{tabular}
\end{table}

\begin{table}[H]
\centering
\caption{Goodness-of-fit, discrete data, case studies of type variance}
\label{tab:GDvariance}

\begin{tabular}{llllllll}
\toprule
\multicolumn{8}{l}{\textbf{Mean Power}} \\
\midrule
  \hline
  \hline
NS-A & NS-B & BJ & AD & K & l-P & W & s-P \\ 
  83 & 80 & 80 & 65 & 62 & 62 & 60 & 54 \\ 
  ----- & ----- & ----- & ----- & ----- & ----- & ----- & ----- \\ 
  Wassp1 & KS & CvM &  &  &  &  &  \\ 
  41 & 30 & 29 &  &  &  &  &  \\ 
   \hline
\addlinespace
\multicolumn{8}{l}{\textbf{Percentage close to best}} \\
\midrule
  \hline
  \hline
NS-A & BJ & NS-B & W & K & AD & Wassp1 & l-P \\ 
  75 & 65 & 60 & 25 & 15 & 10 & 10 & 5 \\ 
  ----- & ----- & ----- & ----- & ----- & ----- & ----- & ----- \\ 
  s-P & KS & CvM &  &  &  &  &  \\ 
  5 & 0 & 0 &  &  &  &  &  \\ 
   \hline
\addlinespace
\multicolumn{8}{l}{\textbf{Mean shortfall from best}} \\
\midrule
  \hline
  \hline
NS-A & NS-B & BJ & AD & K & l-P & W & s-P \\ 
  5 & 8 & 8 & 22 & 25 & 26 & 27 & 34 \\ 
  ----- & ----- & ----- & ----- & ----- & ----- & ----- & ----- \\ 
  Wassp1 & KS & CvM &  &  &  &  &  \\ 
  47 & 57 & 59 &  &  &  &  &  \\ 
   \hline
\addlinespace
\multicolumn{8}{l}{\textbf{Best selection(s)}} \\
\midrule
  \hline
  \hline
NS-A & NS-B \\ 
   \hline
\bottomrule
\end{tabular}
\end{table}

\begin{table}[H]
\centering
\caption{Goodness-of-fit, discrete data, case studies of type skewness}
\label{tab:GDskewness}

\begin{tabular}{llllllll}
\toprule
\multicolumn{8}{l}{\textbf{Mean Power}} \\
\midrule
  \hline
  \hline
NS-A & NS-B & W & s-P & K & l-P & BJ & AD \\ 
  82 & 77 & 75 & 74 & 74 & 64 & 61 & 48 \\ 
  ----- & ----- & ----- & ----- & ----- & ----- & ----- & ----- \\ 
  KS & CvM & Wassp1 &  &  &  &  &  \\ 
  38 & 35 & 30 &  &  &  &  &  \\ 
   \hline
\addlinespace
\multicolumn{8}{l}{\textbf{Percentage close to best}} \\
\midrule
  \hline
  \hline
NS-A & s-P & NS-B & W & K & BJ & l-P & KS \\ 
  90 & 60 & 60 & 40 & 35 & 10 & 5 & 0 \\ 
  ----- & ----- & ----- & ----- & ----- & ----- & ----- & ----- \\ 
  AD & CvM & Wassp1 &  &  &  &  &  \\ 
  0 & 0 & 0 &  &  &  &  &  \\ 
   \hline
\addlinespace
\multicolumn{8}{l}{\textbf{Mean shortfall from best}} \\
\midrule
  \hline
  \hline
NS-A & NS-B & W & s-P & K & l-P & BJ & AD \\ 
  3 & 9 & 10 & 12 & 12 & 21 & 25 & 38 \\ 
  ----- & ----- & ----- & ----- & ----- & ----- & ----- & ----- \\ 
  KS & CvM & Wassp1 &  &  &  &  &  \\ 
  47 & 51 & 56 &  &  &  &  &  \\ 
   \hline
\addlinespace
\multicolumn{8}{l}{\textbf{Best selection(s)}} \\
\midrule
  \hline
  \hline
K & NS-A \\ 
  NS-A & W \\ 
   \hline
\bottomrule
\end{tabular}
\end{table}

\begin{table}[H]
\centering
\caption{Goodness-of-fit, discrete data, case studies of type kurtosis}
\label{tab:GDkurtosis}

\begin{tabular}{llllllll}
\toprule
\multicolumn{8}{l}{\textbf{Mean Power}} \\
\midrule
  \hline
  \hline
NS-A & NS-B & BJ & l-P & K & s-P & W & AD \\ 
  90 & 74 & 72 & 70 & 66 & 62 & 62 & 49 \\ 
  ----- & ----- & ----- & ----- & ----- & ----- & ----- & ----- \\ 
  KS & Wassp1 & CvM &  &  &  &  &  \\ 
  32 & 28 & 26 &  &  &  &  &  \\ 
   \hline
\addlinespace
\multicolumn{8}{l}{\textbf{Percentage close to best}} \\
\midrule
  \hline
  \hline
NS-A & K & W & BJ & s-P & l-P & NS-B & KS \\ 
  85 & 30 & 30 & 30 & 15 & 10 & 10 & 0 \\ 
  ----- & ----- & ----- & ----- & ----- & ----- & ----- & ----- \\ 
  AD & CvM & Wassp1 &  &  &  &  &  \\ 
  0 & 0 & 0 &  &  &  &  &  \\ 
   \hline
\addlinespace
\multicolumn{8}{l}{\textbf{Mean shortfall from best}} \\
\midrule
  \hline
  \hline
NS-A & NS-B & BJ & l-P & K & s-P & W & AD \\ 
  3 & 19 & 20 & 22 & 27 & 30 & 30 & 43 \\ 
  ----- & ----- & ----- & ----- & ----- & ----- & ----- & ----- \\ 
  KS & Wassp1 & CvM &  &  &  &  &  \\ 
  60 & 65 & 66 &  &  &  &  &  \\ 
   \hline
\addlinespace
\multicolumn{8}{l}{\textbf{Best selection(s)}} \\
\midrule
  \hline
  \hline
NS-A & W \\ 
  K & NS-A \\ 
   \hline
\bottomrule
\end{tabular}
\end{table}

\begin{table}[H]
\centering
\caption{Goodness-of-fit, discrete data, case studies of type mixed}
\label{tab:GDmixed}

\begin{tabular}{llllllll}
\toprule
\multicolumn{8}{l}{\textbf{Mean Power}} \\
\midrule
  \hline
  \hline
NS-B & AD & W & Wassp1 & BJ & K & CvM & KS \\ 
  64 & 61 & 60 & 59 & 59 & 59 & 53 & 50 \\ 
  ----- & ----- & ----- & ----- & ----- & ----- & ----- & ----- \\ 
  NS-A & s-P & l-P &  &  &  &  &  \\ 
  50 & 47 & 35 &  &  &  &  &  \\ 
   \hline
\addlinespace
\multicolumn{8}{l}{\textbf{Percentage close to best}} \\
\midrule
  \hline
  \hline
NS-B & AD & CvM & Wassp1 & W & K & BJ & NS-A \\ 
  50 & 48 & 42 & 42 & 40 & 25 & 22 & 20 \\ 
  ----- & ----- & ----- & ----- & ----- & ----- & ----- & ----- \\ 
  KS & s-P & l-P &  &  &  &  &  \\ 
  12 & 12 & 2 &  &  &  &  &  \\ 
   \hline
\addlinespace
\multicolumn{8}{l}{\textbf{Mean shortfall from best}} \\
\midrule
  \hline
  \hline
NS-B & AD & W & Wassp1 & BJ & K & CvM & KS \\ 
  15 & 17 & 18 & 20 & 20 & 20 & 26 & 29 \\ 
  ----- & ----- & ----- & ----- & ----- & ----- & ----- & ----- \\ 
  NS-A & s-P & l-P &  &  &  &  &  \\ 
  29 & 32 & 43 &  &  &  &  &  \\ 
   \hline
\addlinespace
\multicolumn{8}{l}{\textbf{Best selection(s)}} \\
\midrule
  \hline
  \hline
BJ & K & NS-A & NS-B & Wassp1 \\ 
  BJ & NS-A & NS-B & W & Wassp1 \\ 
   \hline
\bottomrule
\end{tabular}
\end{table}

\begin{table}[H]
\centering
\caption{Goodness-of-fit, discrete data, case studies of type mixed without parameter estimation}
\label{tab:GDnE}

\begin{tabular}{llllllll}
\toprule
\multicolumn{8}{l}{\textbf{Mean Power}} \\
\midrule
  \hline
  \hline
NS-B & BJ & AD & Wassp1 & W & K & NS-A & s-P \\ 
  68 & 64 & 59 & 58 & 56 & 56 & 56 & 51 \\ 
  ----- & ----- & ----- & ----- & ----- & ----- & ----- & ----- \\ 
  CvM & KS & l-P &  &  &  &  &  \\ 
  45 & 44 & 42 &  &  &  &  &  \\ 
   \hline
\addlinespace
\multicolumn{8}{l}{\textbf{Percentage close to best}} \\
\midrule
  \hline
  \hline
NS-B & W & Wassp1 & AD & CvM & NS-A & BJ & K \\ 
  50 & 40 & 35 & 30 & 25 & 25 & 20 & 15 \\ 
  ----- & ----- & ----- & ----- & ----- & ----- & ----- & ----- \\ 
  s-P & KS & l-P &  &  &  &  &  \\ 
  15 & 10 & 5 &  &  &  &  &  \\ 
   \hline
\addlinespace
\multicolumn{8}{l}{\textbf{Mean shortfall from best}} \\
\midrule
  \hline
  \hline
NS-B & BJ & AD & Wassp1 & W & K & NS-A & s-P \\ 
  15 & 18 & 24 & 25 & 27 & 27 & 27 & 32 \\ 
  ----- & ----- & ----- & ----- & ----- & ----- & ----- & ----- \\ 
  CvM & KS & l-P &  &  &  &  &  \\ 
  37 & 39 & 41 &  &  &  &  &  \\ 
   \hline
\addlinespace
\multicolumn{8}{l}{\textbf{Best selection(s)}} \\
\midrule
  \hline
  \hline
BJ & NS-A & W & Wassp1 \\ 
  BJ & NS-A & NS-B & Wassp1 \\ 
   \hline
\bottomrule
\end{tabular}
\end{table}

\begin{table}[H]
\centering
\caption{Goodness-of-fit, discrete data, case studies of type mixed with parameter estimation}
\label{tab:GDE}

\begin{tabular}{llllllll}
\toprule
\multicolumn{8}{l}{\textbf{Mean Power}} \\
\midrule
  \hline
  \hline
W & AD & K & CvM & NS-B & Wassp1 & KS & BJ \\ 
  64 & 63 & 61 & 60 & 60 & 59 & 56 & 53 \\ 
  ----- & ----- & ----- & ----- & ----- & ----- & ----- & ----- \\ 
  NS-A & s-P & l-P &  &  &  &  &  \\ 
  44 & 42 & 28 &  &  &  &  &  \\ 
   \hline
\addlinespace
\multicolumn{8}{l}{\textbf{Percentage close to best}} \\
\midrule
  \hline
  \hline
AD & CvM & Wassp1 & NS-B & W & K & BJ & KS \\ 
  65 & 60 & 50 & 50 & 40 & 35 & 25 & 15 \\ 
  ----- & ----- & ----- & ----- & ----- & ----- & ----- & ----- \\ 
  NS-A & s-P & l-P &  &  &  &  &  \\ 
  15 & 10 & 0 &  &  &  &  &  \\ 
   \hline
\addlinespace
\multicolumn{8}{l}{\textbf{Mean shortfall from best}} \\
\midrule
  \hline
  \hline
W & AD & K & CvM & NS-B & Wassp1 & KS & BJ \\ 
  10 & 11 & 13 & 14 & 15 & 15 & 19 & 21 \\ 
  ----- & ----- & ----- & ----- & ----- & ----- & ----- & ----- \\ 
  NS-A & s-P & l-P &  &  &  &  &  \\ 
  30 & 32 & 46 &  &  &  &  &  \\ 
   \hline
\addlinespace
\multicolumn{8}{l}{\textbf{Best selection(s)}} \\
\midrule
  \hline
  \hline
BJ & K & NS-A & NS-B \\ 
  BJ & NS-A & NS-B & W \\ 
   \hline
\bottomrule
\end{tabular}
\end{table}

The type of the difference matters here as well (Tables \ref{tab:GDmean} to \ref{tab:GDE}). NS-A has the highest mean power if the distributions differ in the variance, the skewness or the kurtosis, Anderson-Darling if they differ in the mean. In the case studies of type mixed NS-B is best without parameter estimation, but with parameter estimation Watson, Anderson-Darling and Kuiper are best, and NS-A falls far behind.

\subsection{Goodness-of-fit plus hybrid methods}

To distinguish between methods that exist for both the goodness-of-fit and the two-sample problem the abbreviations of the first end in G and of the second in T1 if the MC sample size is equal to the one of the data set and in T5 if it is five times as large. Only methods with a mean power of at least $40\%$ are included, otherwise there are too many methods for the search of the best selections.

\begin{table}[H]
\centering
\caption{Power of goodness-of-fit and hybrid methods, continuous data, all case studies, methods with a mean power of at least 40\% only}
\label{tab:GHcont}

\begin{tabular}{lllllllll}
\toprule
\multicolumn{9}{l}{\textbf{Mean Power}} \\
\midrule
  \hline
  \hline
ZCG & ZAG & ZKG & ZCT5 & ZAT5 & HHGT5 & ADG & WG & ZKT5 \\ 
  69 & 66 & 62 & 60 & 60 & 53 & 51 & 50 & 50 \\ 
  ----- & ----- & ----- & ----- & ----- & ----- & ----- & ----- & ----- \\ 
  KG & EppsST5 & Wassp1G & ADT5 & ES-s-LG & EP-s-LG & ES-s-PG & KuiperT5 & EP-s-PG \\ 
  49 & 48 & 45 & 45 & 44 & 43 & 43 & 43 & 43 \\ 
  ----- & ----- & ----- & ----- & ----- & ----- & ----- & ----- & ----- \\ 
  CvMG & KSG &  &  &  &  &  &  &  \\ 
  42 & 40 &  &  &  &  &  &  &  \\ 
   \hline
\addlinespace
\multicolumn{9}{l}{\textbf{Percentage close to best}} \\
\midrule
  \hline
  \hline
ZCG & ZAG & ADG & ZCT5 & Wassp1G & WG & ZAT5 & CvMG & KG \\ 
  58 & 45 & 28 & 28 & 26 & 23 & 22 & 21 & 18 \\ 
  ----- & ----- & ----- & ----- & ----- & ----- & ----- & ----- & ----- \\ 
  EppsST5 & ZKG & ES-s-LG & KSG & ES-s-PG & EP-s-LG & EP-s-PG & ADT5 & HHGT5 \\ 
  18 & 16 & 8 & 8 & 8 & 8 & 7 & 7 & 7 \\ 
  ----- & ----- & ----- & ----- & ----- & ----- & ----- & ----- & ----- \\ 
  KuiperT5 & ZKT5 &  &  &  &  &  &  &  \\ 
  4 & 0 &  &  &  &  &  &  &  \\ 
   \hline
\addlinespace
\multicolumn{9}{l}{\textbf{Mean shortfall from best}} \\
\midrule
  \hline
  \hline
ZCG & ZAG & ZKG & ZCT5 & ZAT5 & HHGT5 & ADG & WG & ZKT5 \\ 
  12 & 14 & 18 & 20 & 21 & 27 & 29 & 30 & 31 \\ 
  ----- & ----- & ----- & ----- & ----- & ----- & ----- & ----- & ----- \\ 
  KG & EppsST5 & Wassp1G & ADT5 & ES-s-LG & EP-s-LG & ES-s-PG & KuiperT5 & EP-s-PG \\ 
  31 & 32 & 35 & 36 & 37 & 37 & 37 & 38 & 38 \\ 
  ----- & ----- & ----- & ----- & ----- & ----- & ----- & ----- & ----- \\ 
  CvMG & KSG &  &  &  &  &  &  &  \\ 
  38 & 41 &  &  &  &  &  &  &  \\ 
   \hline
\addlinespace
\multicolumn{9}{l}{\textbf{Best selection(s)}} \\
\midrule
  \hline
  \hline
EppsST5 & HHGT5 & Wassp1G & WG & ZAG & ZCG \\ 
   \hline
\bottomrule
\end{tabular}
\end{table}

\begin{figure}[H]
\centering
\includegraphics[width=0.85\textwidth]{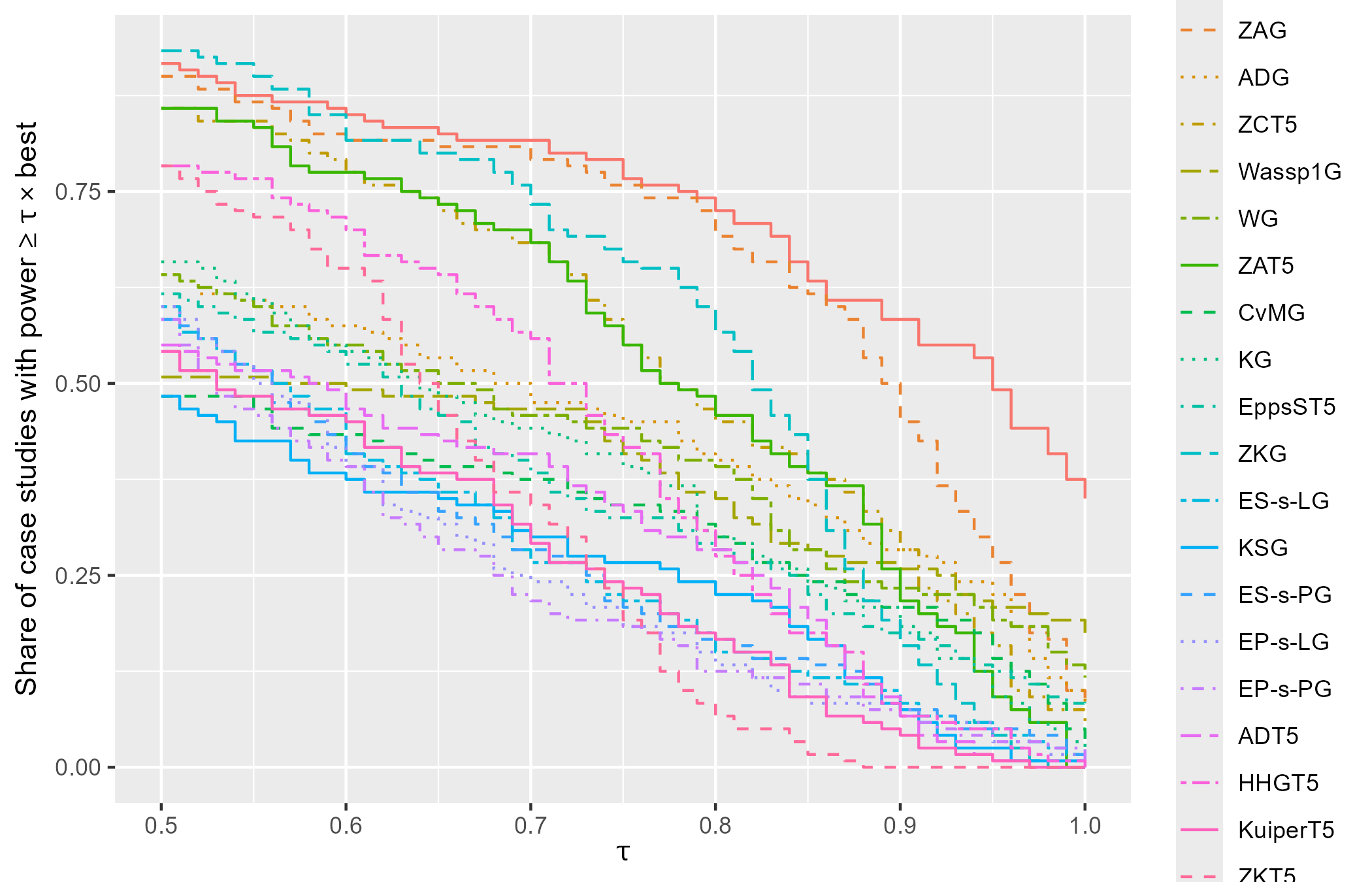}
\caption{Performance profiles, goodness-of-fit and hybrid methods, continuous data.}
\label{fig:profGHcont}
\end{figure}

\begin{table}[H]
\centering
\caption{Power of goodness-of-fit and hybrid methods, discrete data, all case studies, methods with a mean power of at least 40\% only}
\label{tab:GHdisc}

\begin{tabular}{lllllllll}
\toprule
\multicolumn{9}{l}{\textbf{Mean Power}} \\
\midrule
  \hline
  \hline
NS-BG & NS-AG & BJG & HHGT5 & KG & ADG & WG & EppsST5 & s-PG \\ 
  71 & 68 & 66 & 64 & 60 & 59 & 59 & 58 & 56 \\ 
  ----- & ----- & ----- & ----- & ----- & ----- & ----- & ----- & ----- \\ 
  l-PG & ADT5 & KuiperT5 & smallT5 & EppsST1 & Wassp1G & largeT5 & CvMG & KSG \\ 
  52 & 52 & 51 & 48 & 46 & 46 & 44 & 43 & 43 \\ 
  ----- & ----- & ----- & ----- & ----- & ----- & ----- & ----- & ----- \\ 
  HHGT1 &  &  &  &  &  &  &  &  \\ 
  40 &  &  &  &  &  &  &  &  \\ 
   \hline
\addlinespace
\multicolumn{9}{l}{\textbf{Percentage close to best}} \\
\midrule
  \hline
  \hline
NS-AG & NS-BG & ADG & BJG & WG & EppsST5 & HHGT5 & KG & CvMG \\ 
  54 & 39 & 31 & 31 & 30 & 23 & 23 & 22 & 20 \\ 
  ----- & ----- & ----- & ----- & ----- & ----- & ----- & ----- & ----- \\ 
  Wassp1G & s-PG & ADT5 & KSG & KuiperT5 & l-PG & smallT5 & EppsST1 & largeT5 \\ 
  20 & 18 & 12 & 8 & 4 & 3 & 2 & 2 & 1 \\ 
  ----- & ----- & ----- & ----- & ----- & ----- & ----- & ----- & ----- \\ 
  HHGT1 &  &  &  &  &  &  &  &  \\ 
  0 &  &  &  &  &  &  &  &  \\ 
   \hline
\addlinespace
\multicolumn{9}{l}{\textbf{Mean shortfall from best}} \\
\midrule
  \hline
  \hline
NS-BG & NS-AG & BJG & HHGT5 & KG & ADG & WG & EppsST5 & s-PG \\ 
  13 & 16 & 18 & 20 & 24 & 24 & 25 & 25 & 28 \\ 
  ----- & ----- & ----- & ----- & ----- & ----- & ----- & ----- & ----- \\ 
  l-PG & ADT5 & KuiperT5 & smallT5 & EppsST1 & Wassp1G & largeT5 & CvMG & KSG \\ 
  31 & 32 & 32 & 36 & 38 & 38 & 39 & 41 & 41 \\ 
  ----- & ----- & ----- & ----- & ----- & ----- & ----- & ----- & ----- \\ 
  HHGT1 &  &  &  &  &  &  &  &  \\ 
  44 &  &  &  &  &  &  &  &  \\ 
   \hline
\addlinespace
\multicolumn{9}{l}{\textbf{Best selection(s)}} \\
\midrule
  \hline
  \hline
ADG & BJG & EppsST5 & NS-AG & NS-BG & Wassp1G & WG \\ 
  ADG & BJG & EppsST5 & KG & NS-AG & NS-BG & Wassp1G \\ 
   \hline
\bottomrule
\end{tabular}
\end{table}

\begin{figure}[H]
\centering
\includegraphics[width=0.85\textwidth]{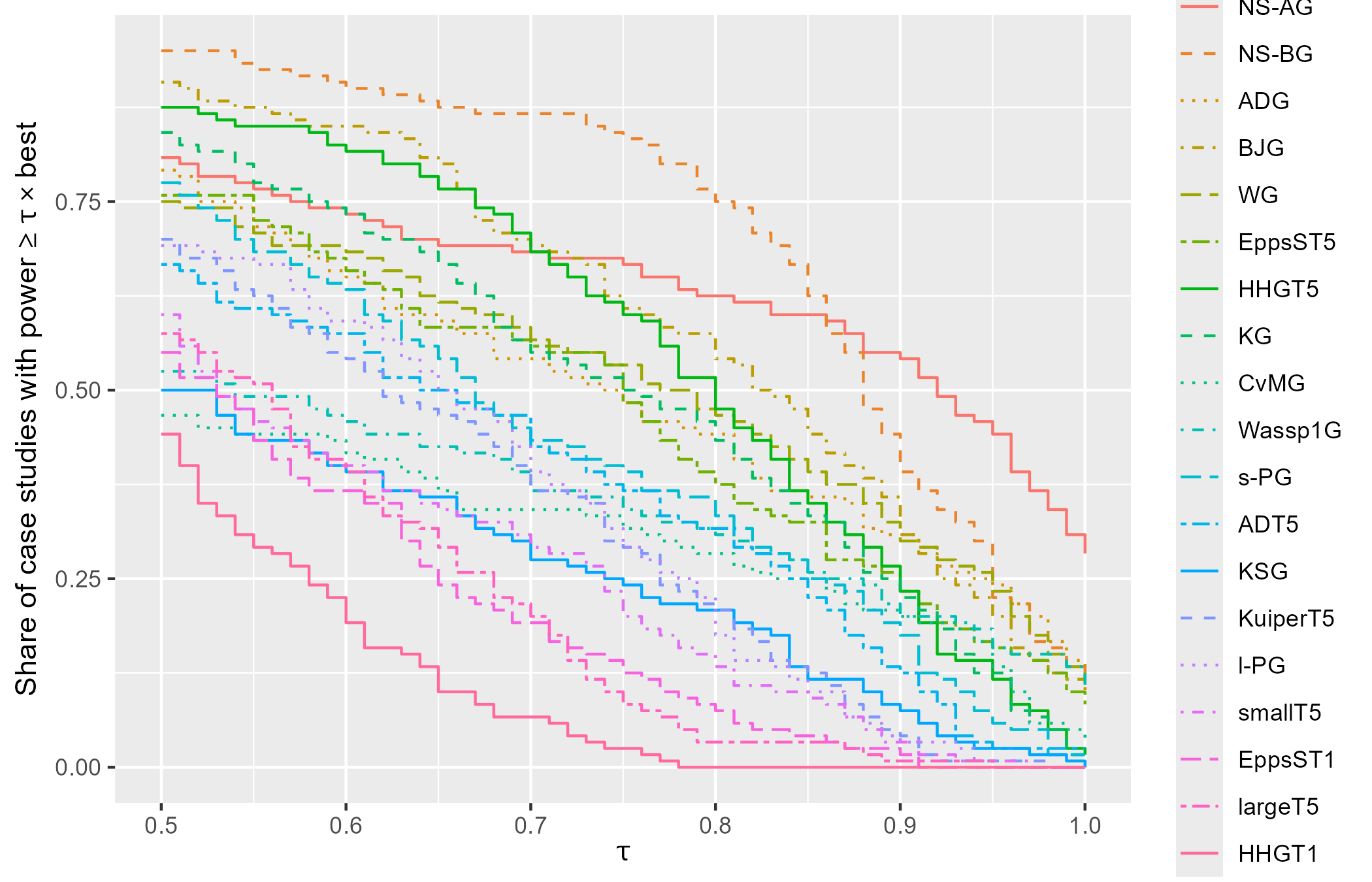}
\caption{Performance profiles, goodness-of-fit and hybrid methods, discrete data.}
\label{fig:profGHdisc}
\end{figure}

A larger MC data set clearly helps: for continuous data the mean power of the hybrid ZC test increases from $36\%$ with nMC=1 (not in table \ref{tab:GHcont}, which shows only methods with a mean power of at least $40\%$) to $60\%$ with nMC=5. But even then it does not reach the power of ZC as a goodness-of-fit test ($69\%$). For continuous data the hybrid HHG and Epps-Singleton tests with nMC=5 are part of the best selection, together with the goodness-of-fit tests ZC, ZA, Watson and Wasserstein, but they are within $90\%$ of the best method in only $7\%$ and $18\%$ of the case studies. For discrete data the best hybrid methods with nMC=5, HHG and Epps-Singleton, have a mean power of $64\%$ and $58\%$, compared to $71\%$ for the best goodness-of-fit test, NS-B. Here the hybrid Epps-Singleton test with nMC=5 is part of the best selection, together with the goodness-of-fit tests NS-B, NS-A, Berk-Jones, Anderson-Darling, Wasserstein and Watson (or Kuiper). It is within $90\%$ of the best method in $23\%$ of the case studies.

\subsection{Two-sample}

\subsubsection{Continuous data, all case studies}

\begin{table}[H]
\centering
\caption{Two-sample, continuous data, all case studies}
\label{tab:TC}

\begin{tabular}{lllllllll}
\toprule
\multicolumn{9}{l}{\textbf{Mean Power}} \\
\midrule
  \hline
  \hline
ZC & ZA & EppsS & HHG & ZK & ES small & EP small & Kuiper & AD \\ 
  73 & 73 & 67 & 63 & 63 & 49 & 49 & 47 & 43 \\ 
  ----- & ----- & ----- & ----- & ----- & ----- & ----- & ----- & ----- \\ 
  ES large & EP large & Wassp1 & CvM & KS &  &  &  &  \\ 
  38 & 33 & 33 & 32 & 31 &  &  &  &  \\ 
   \hline
\addlinespace
\multicolumn{9}{l}{\textbf{Percentage close to best}} \\
\midrule
  \hline
  \hline
ZA & ZC & EppsS & HHG & Kuiper & AD & ZK & CvM & Wassp1 \\ 
  67 & 67 & 50 & 30 & 16 & 14 & 14 & 11 & 11 \\ 
  ----- & ----- & ----- & ----- & ----- & ----- & ----- & ----- & ----- \\ 
  EP small & ES small & KS & ES large & EP large &  &  &  &  \\ 
  6 & 4 & 3 & 0 & 0 &  &  &  &  \\ 
   \hline
\addlinespace
\multicolumn{9}{l}{\textbf{Mean shortfall from best}} \\
\midrule
  \hline
  \hline
ZC & ZA & EppsS & HHG & ZK & ES small & EP small & Kuiper & AD \\ 
  9 & 10 & 15 & 20 & 20 & 34 & 34 & 35 & 39 \\ 
  ----- & ----- & ----- & ----- & ----- & ----- & ----- & ----- & ----- \\ 
  ES large & EP large & Wassp1 & CvM & KS &  &  &  &  \\ 
  44 & 49 & 49 & 51 & 51 &  &  &  &  \\ 
   \hline
\addlinespace
\multicolumn{9}{l}{\textbf{Best selection(s)}} \\
\midrule
  \hline
  \hline
CvM & EppsS & HHG & ZA \\ 
  CvM & EppsS & HHG & ZC \\ 
  EppsS & HHG & Wassp1 & ZA \\ 
  EppsS & HHG & Wassp1 & ZC \\ 
  AD & EppsS & HHG & ZA \\ 
  AD & EppsS & HHG & ZC \\ 
   \hline
\bottomrule
\end{tabular}
\end{table}

\begin{figure}[H]
\centering
\includegraphics[width=0.85\textwidth]{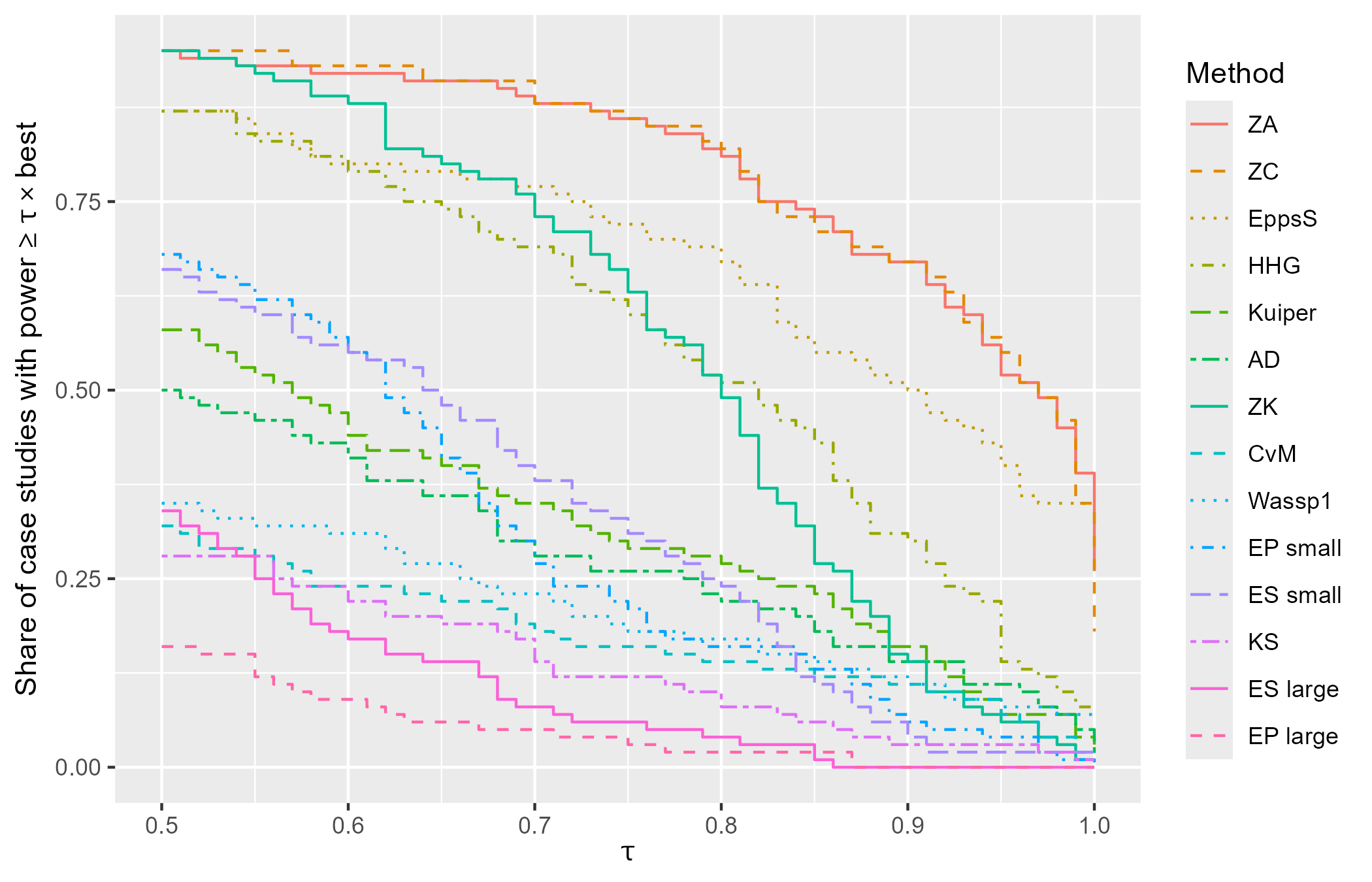}
\caption{Performance profiles, two-sample problem, continuous data.}
\label{fig:profTC}
\end{figure}

Here the Zhang tests ZC and ZA have the highest mean power, followed by the Epps-Singleton and the HHG tests. In the case studies of type variance either one of the Zhang tests alone is close to the best in almost all of them (Tables \ref{tab:TCmean} to \ref{tab:TCmixed}). Four methods are needed overall: ZC (or ZA), Epps-Singleton, HHG and Anderson-Darling (or Cram\'er-von Mises or Wasserstein).

\subsubsection{Continuous data, by type}

\begin{table}[H]
\centering
\caption{Two-sample, continuous data, case studies of type mean}
\label{tab:TCmean}

\begin{tabular}{lllllllll}
\toprule
\multicolumn{9}{l}{\textbf{Mean Power}} \\
\midrule
  \hline
  \hline
ZC & ZA & HHG & ZK & AD & EppsS & CvM & Wassp1 & KS \\ 
  78 & 77 & 69 & 68 & 68 & 64 & 58 & 53 & 52 \\ 
  ----- & ----- & ----- & ----- & ----- & ----- & ----- & ----- & ----- \\ 
  EP small & ES small & Kuiper & ES large & EP large &  &  &  &  \\ 
  51 & 47 & 35 & 32 & 27 &  &  &  &  \\ 
   \hline
\addlinespace
\multicolumn{9}{l}{\textbf{Percentage close to best}} \\
\midrule
  \hline
  \hline
ZA & ZC & HHG & AD & CvM & Wassp1 & EppsS & KS & ZK \\ 
  90 & 90 & 45 & 35 & 30 & 20 & 20 & 10 & 10 \\ 
  ----- & ----- & ----- & ----- & ----- & ----- & ----- & ----- & ----- \\ 
  Kuiper & ES small & ES large & EP large & EP small &  &  &  &  \\ 
  5 & 5 & 0 & 0 & 0 &  &  &  &  \\ 
   \hline
\addlinespace
\multicolumn{9}{l}{\textbf{Mean shortfall from best}} \\
\midrule
  \hline
  \hline
ZC & ZA & HHG & ZK & AD & EppsS & CvM & Wassp1 & KS \\ 
  3 & 3 & 12 & 13 & 13 & 17 & 23 & 28 & 29 \\ 
  ----- & ----- & ----- & ----- & ----- & ----- & ----- & ----- & ----- \\ 
  EP small & ES small & Kuiper & ES large & EP large &  &  &  &  \\ 
  30 & 34 & 46 & 49 & 53 &  &  &  &  \\ 
   \hline
\addlinespace
\multicolumn{9}{l}{\textbf{Best selection(s)}} \\
\midrule
  \hline
  \hline
KS & ZA \\ 
  KS & ZC \\ 
  CvM & ZA \\ 
  CvM & ZC \\ 
  AD & ZA \\ 
  AD & ZC \\ 
  HHG & ZA \\ 
  HHG & ZC \\ 
   \hline
\bottomrule
\end{tabular}
\end{table}

\begin{table}[H]
\centering
\caption{Two-sample, continuous data, case studies of type variance}
\label{tab:TCvariance}

\begin{tabular}{lllllllll}
\toprule
\multicolumn{9}{l}{\textbf{Mean Power}} \\
\midrule
  \hline
  \hline
ZA & ZC & ZK & EppsS & HHG & Kuiper & ES small & EP small & AD \\ 
  80 & 80 & 66 & 62 & 51 & 40 & 37 & 36 & 34 \\ 
  ----- & ----- & ----- & ----- & ----- & ----- & ----- & ----- & ----- \\ 
  ES large & EP large & Wassp1 & CvM & KS &  &  &  &  \\ 
  31 & 28 & 24 & 18 & 17 &  &  &  &  \\ 
   \hline
\addlinespace
\multicolumn{9}{l}{\textbf{Percentage close to best}} \\
\midrule
  \hline
  \hline
ZA & ZC & EppsS & ZK & Kuiper & HHG & Wassp1 & KS & CvM \\ 
  95 & 95 & 45 & 15 & 10 & 10 & 5 & 0 & 0 \\ 
  ----- & ----- & ----- & ----- & ----- & ----- & ----- & ----- & ----- \\ 
  AD & ES large & ES small & EP large & EP small &  &  &  &  \\ 
  0 & 0 & 0 & 0 & 0 &  &  &  &  \\ 
   \hline
\addlinespace
\multicolumn{9}{l}{\textbf{Mean shortfall from best}} \\
\midrule
  \hline
  \hline
ZA & ZC & ZK & EppsS & HHG & Kuiper & ES small & EP small & AD \\ 
  1 & 1 & 15 & 19 & 30 & 41 & 44 & 45 & 47 \\ 
  ----- & ----- & ----- & ----- & ----- & ----- & ----- & ----- & ----- \\ 
  ES large & EP large & Wassp1 & CvM & KS &  &  &  &  \\ 
  50 & 53 & 57 & 63 & 64 &  &  &  &  \\ 
   \hline
\addlinespace
\multicolumn{9}{l}{\textbf{Best selection(s)}} \\
\midrule
  \hline
  \hline
Kuiper & ZC \\ 
  Kuiper & ZA \\ 
  HHG & ZC \\ 
  HHG & ZA \\ 
  EppsS & ZC \\ 
  EppsS & ZA \\ 
   \hline
\bottomrule
\end{tabular}
\end{table}

\begin{table}[H]
\centering
\caption{Two-sample, continuous data, case studies of type skewness}
\label{tab:TCskewness}

\begin{tabular}{lllllllll}
\toprule
\multicolumn{9}{l}{\textbf{Mean Power}} \\
\midrule
  \hline
  \hline
EppsS & HHG & ZC & ZA & EP small & Kuiper & ES small & ZK & ES large \\ 
  75 & 72 & 69 & 68 & 61 & 61 & 58 & 58 & 47 \\ 
  ----- & ----- & ----- & ----- & ----- & ----- & ----- & ----- & ----- \\ 
  EP large & AD & KS & CvM & Wassp1 &  &  &  &  \\ 
  41 & 36 & 28 & 25 & 20 &  &  &  &  \\ 
   \hline
\addlinespace
\multicolumn{9}{l}{\textbf{Percentage close to best}} \\
\midrule
  \hline
  \hline
EppsS & HHG & ZA & ZC & Kuiper & ZK & EP small & ES small & KS \\ 
  70 & 40 & 35 & 35 & 20 & 15 & 10 & 5 & 0 \\ 
  ----- & ----- & ----- & ----- & ----- & ----- & ----- & ----- & ----- \\ 
  CvM & AD & Wassp1 & ES large & EP large &  &  &  &  \\ 
  0 & 0 & 0 & 0 & 0 &  &  &  &  \\ 
   \hline
\addlinespace
\multicolumn{9}{l}{\textbf{Mean shortfall from best}} \\
\midrule
  \hline
  \hline
EppsS & HHG & ZC & ZA & EP small & Kuiper & ES small & ZK & ES large \\ 
  10 & 13 & 15 & 16 & 23 & 24 & 26 & 27 & 38 \\ 
  ----- & ----- & ----- & ----- & ----- & ----- & ----- & ----- & ----- \\ 
  EP large & AD & KS & CvM & Wassp1 &  &  &  &  \\ 
  44 & 48 & 56 & 60 & 65 &  &  &  &  \\ 
   \hline
\addlinespace
\multicolumn{9}{l}{\textbf{Best selection(s)}} \\
\midrule
  \hline
  \hline
EppsS & HHG & ZA \\ 
  EppsS & HHG & ZC \\ 
   \hline
\bottomrule
\end{tabular}
\end{table}

\begin{table}[H]
\centering
\caption{Two-sample, continuous data, case studies of type kurtosis}
\label{tab:TCkurtosis}

\begin{tabular}{lllllllll}
\toprule
\multicolumn{9}{l}{\textbf{Mean Power}} \\
\midrule
  \hline
  \hline
ZA & ZC & EppsS & ZK & HHG & ES small & ES large & Kuiper & EP small \\ 
  72 & 72 & 67 & 62 & 57 & 45 & 45 & 44 & 41 \\ 
  ----- & ----- & ----- & ----- & ----- & ----- & ----- & ----- & ----- \\ 
  EP large & AD & KS & Wassp1 & CvM &  &  &  &  \\ 
  41 & 24 & 19 & 15 & 15 &  &  &  &  \\ 
   \hline
\addlinespace
\multicolumn{9}{l}{\textbf{Percentage close to best}} \\
\midrule
  \hline
  \hline
EppsS & ZA & ZC & ZK & Kuiper & HHG & EP small & KS & CvM \\ 
  65 & 50 & 50 & 30 & 15 & 10 & 5 & 0 & 0 \\ 
  ----- & ----- & ----- & ----- & ----- & ----- & ----- & ----- & ----- \\ 
  AD & Wassp1 & ES large & ES small & EP large &  &  &  &  \\ 
  0 & 0 & 0 & 0 & 0 &  &  &  &  \\ 
   \hline
\addlinespace
\multicolumn{9}{l}{\textbf{Mean shortfall from best}} \\
\midrule
  \hline
  \hline
ZA & ZC & EppsS & ZK & HHG & ES small & ES large & Kuiper & EP small \\ 
  13 & 14 & 19 & 23 & 29 & 40 & 40 & 41 & 44 \\ 
  ----- & ----- & ----- & ----- & ----- & ----- & ----- & ----- & ----- \\ 
  EP large & AD & KS & Wassp1 & CvM &  &  &  &  \\ 
  45 & 62 & 67 & 71 & 71 &  &  &  &  \\ 
   \hline
\addlinespace
\multicolumn{9}{l}{\textbf{Best selection(s)}} \\
\midrule
  \hline
  \hline
EP small & EppsS & ZC \\ 
  EP small & EppsS & ZA \\ 
  EppsS & Kuiper & ZC \\ 
  EppsS & Kuiper & ZA \\ 
  EppsS & HHG & ZC \\ 
  EppsS & HHG & ZA \\ 
   \hline
\bottomrule
\end{tabular}
\end{table}

\begin{table}[H]
\centering
\caption{Two-sample, continuous data, case studies of type mixed}
\label{tab:TCmixed}

\begin{tabular}{lllllllll}
\toprule
\multicolumn{9}{l}{\textbf{Mean Power}} \\
\midrule
  \hline
  \hline
EppsS & ZC & ZA & HHG & ZK & Kuiper & ES small & Wassp1 & AD \\ 
  68 & 66 & 65 & 65 & 59 & 56 & 56 & 54 & 53 \\ 
  ----- & ----- & ----- & ----- & ----- & ----- & ----- & ----- & ----- \\ 
  EP small & CvM & KS & ES large & EP large &  &  &  &  \\ 
  53 & 43 & 41 & 36 & 31 &  &  &  &  \\ 
   \hline
\addlinespace
\multicolumn{9}{l}{\textbf{Percentage close to best}} \\
\midrule
  \hline
  \hline
ZA & ZC & EppsS & HHG & AD & Kuiper & Wassp1 & CvM & EP small \\ 
  65 & 65 & 50 & 45 & 35 & 30 & 30 & 25 & 15 \\ 
  ----- & ----- & ----- & ----- & ----- & ----- & ----- & ----- & ----- \\ 
  ES small & KS & ZK & ES large & EP large &  &  &  &  \\ 
  10 & 5 & 0 & 0 & 0 &  &  &  &  \\ 
   \hline
\addlinespace
\multicolumn{9}{l}{\textbf{Mean shortfall from best}} \\
\midrule
  \hline
  \hline
EppsS & ZC & ZA & HHG & ZK & Kuiper & ES small & Wassp1 & AD \\ 
  12 & 13 & 14 & 14 & 21 & 23 & 23 & 25 & 26 \\ 
  ----- & ----- & ----- & ----- & ----- & ----- & ----- & ----- & ----- \\ 
  EP small & CvM & KS & ES large & EP large &  &  &  &  \\ 
  26 & 36 & 38 & 43 & 48 &  &  &  &  \\ 
   \hline
\addlinespace
\multicolumn{9}{l}{\textbf{Best selection(s)}} \\
\midrule
  \hline
  \hline
CvM & HHG & ZA \\ 
  AD & HHG & ZA \\ 
  HHG & Wassp1 & ZA \\ 
   \hline
\bottomrule
\end{tabular}
\end{table}

\subsubsection{Discrete data, all case studies}

\begin{table}[H]
\centering
\caption{Two-sample, discrete data, all case studies}
\label{tab:TD}

\begin{tabular}{llllllll}
\toprule
\multicolumn{8}{l}{\textbf{Mean Power}} \\
\midrule
  \hline
  \hline
EppsS & HHG & AD & small & Kuiper & large & Wassp1 & CvM \\ 
  83 & 83 & 68 & 64 & 62 & 53 & 52 & 47 \\ 
  ----- & ----- & ----- & ----- & ----- & ----- & ----- & ----- \\ 
  KS & ZA &  &  &  &  &  &  \\ 
  46 & 28 &  &  &  &  &  &  \\ 
   \hline
\addlinespace
\multicolumn{8}{l}{\textbf{Percentage close to best}} \\
\midrule
  \hline
  \hline
HHG & EppsS & AD & Wassp1 & Kuiper & CvM & KS & small \\ 
  81 & 71 & 40 & 21 & 18 & 17 & 5 & 5 \\ 
  ----- & ----- & ----- & ----- & ----- & ----- & ----- & ----- \\ 
  ZA & large &  &  &  &  &  &  \\ 
  4 & 4 &  &  &  &  &  &  \\ 
   \hline
\addlinespace
\multicolumn{8}{l}{\textbf{Mean shortfall from best}} \\
\midrule
  \hline
  \hline
EppsS & HHG & AD & small & Kuiper & large & Wassp1 & CvM \\ 
  6 & 6 & 21 & 25 & 27 & 36 & 37 & 42 \\ 
  ----- & ----- & ----- & ----- & ----- & ----- & ----- & ----- \\ 
  KS & ZA &  &  &  &  &  &  \\ 
  43 & 61 &  &  &  &  &  &  \\ 
   \hline
\addlinespace
\multicolumn{8}{l}{\textbf{Best selection(s)}} \\
\midrule
  \hline
  \hline
AD & HHG & Kuiper \\ 
   \hline
\bottomrule
\end{tabular}
\end{table}

\begin{figure}[H]
\centering
\includegraphics[width=0.85\textwidth]{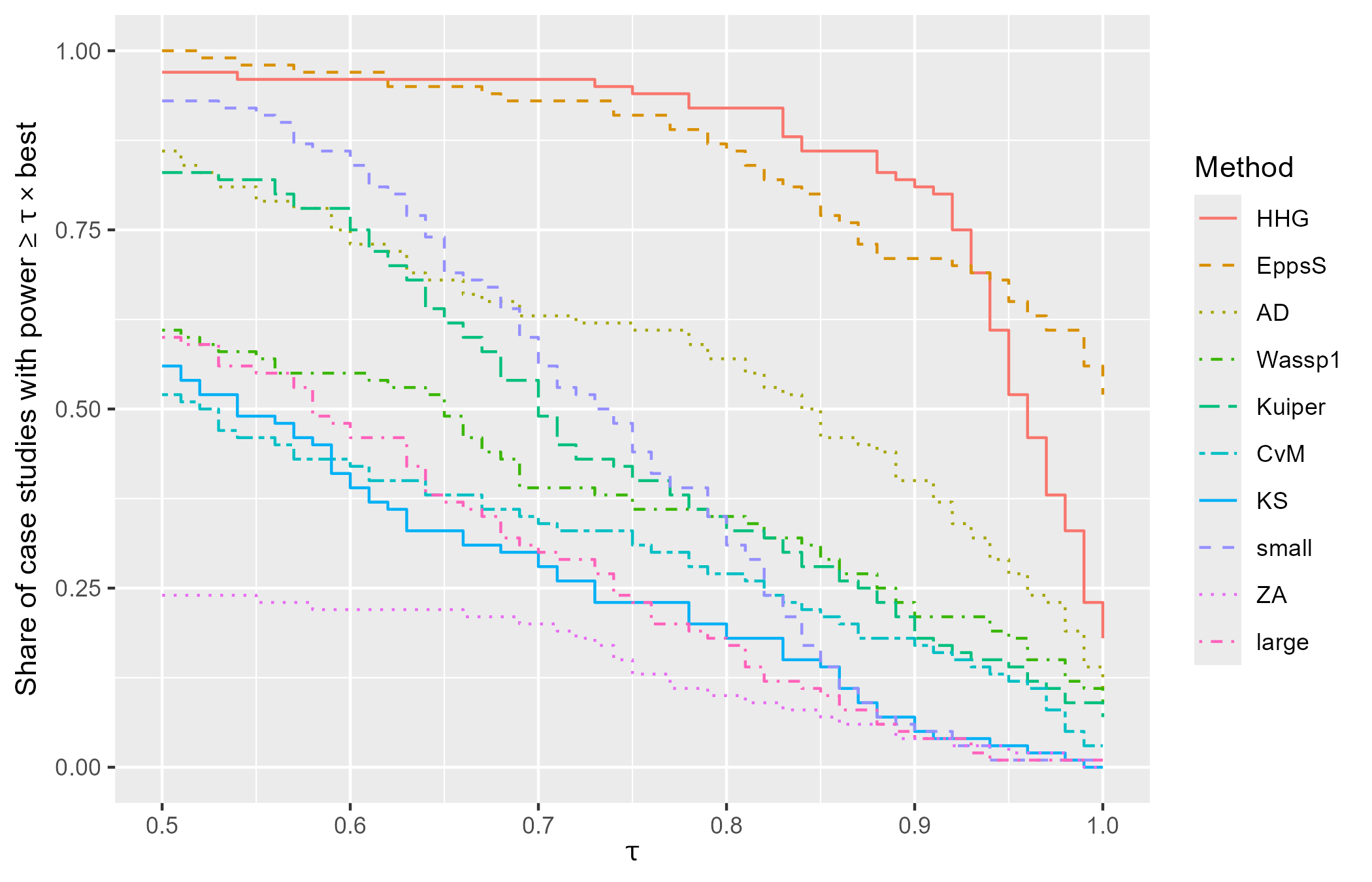}
\caption{Performance profiles, two-sample problem, discrete data.}
\label{fig:profTD}
\end{figure}

For discrete data the Epps-Singleton and the HHG tests have by far the highest mean power, and the discrete version of ZA is weak. The best selection consists of only three methods: HHG, Anderson-Darling and Kuiper. The results by type are in Tables \ref{tab:TDmean} to \ref{tab:TDmixed}.

\subsubsection{Discrete data, by type}

\begin{table}[H]
\centering
\caption{Two-sample, discrete data, case studies of type mean}
\label{tab:TDmean}

\begin{tabular}{llllllll}
\toprule
\multicolumn{8}{l}{\textbf{Mean Power}} \\
\midrule
  \hline
  \hline
AD & HHG & EppsS & CvM & Wassp1 & KS & small & Kuiper \\ 
  81 & 81 & 79 & 72 & 71 & 64 & 63 & 47 \\ 
  ----- & ----- & ----- & ----- & ----- & ----- & ----- & ----- \\ 
  large & ZA &  &  &  &  &  &  \\ 
  41 & 30 &  &  &  &  &  &  \\ 
   \hline
\addlinespace
\multicolumn{8}{l}{\textbf{Percentage close to best}} \\
\midrule
  \hline
  \hline
AD & HHG & EppsS & CvM & Wassp1 & KS & Kuiper & small \\ 
  75 & 75 & 55 & 35 & 35 & 15 & 5 & 5 \\ 
  ----- & ----- & ----- & ----- & ----- & ----- & ----- & ----- \\ 
  ZA & large &  &  &  &  &  &  \\ 
  0 & 0 &  &  &  &  &  &  \\ 
   \hline
\addlinespace
\multicolumn{8}{l}{\textbf{Mean shortfall from best}} \\
\midrule
  \hline
  \hline
AD & HHG & EppsS & CvM & Wassp1 & KS & small & Kuiper \\ 
  5 & 6 & 8 & 15 & 16 & 23 & 24 & 40 \\ 
  ----- & ----- & ----- & ----- & ----- & ----- & ----- & ----- \\ 
  large & ZA &  &  &  &  &  &  \\ 
  45 & 57 &  &  &  &  &  &  \\ 
   \hline
\addlinespace
\multicolumn{8}{l}{\textbf{Best selection(s)}} \\
\midrule
  \hline
  \hline
AD & HHG \\ 
   \hline
\bottomrule
\end{tabular}
\end{table}

\begin{table}[H]
\centering
\caption{Two-sample, discrete data, case studies of type variance}
\label{tab:TDvariance}

\begin{tabular}{llllllll}
\toprule
\multicolumn{8}{l}{\textbf{Mean Power}} \\
\midrule
  \hline
  \hline
EppsS & HHG & AD & Kuiper & small & large & Wassp1 & CvM \\ 
  85 & 82 & 74 & 65 & 62 & 57 & 45 & 37 \\ 
  ----- & ----- & ----- & ----- & ----- & ----- & ----- & ----- \\ 
  KS & ZA &  &  &  &  &  &  \\ 
  35 & 27 &  &  &  &  &  &  \\ 
   \hline
\addlinespace
\multicolumn{8}{l}{\textbf{Percentage close to best}} \\
\midrule
  \hline
  \hline
HHG & EppsS & AD & Kuiper & Wassp1 & CvM & KS & ZA \\ 
  80 & 75 & 40 & 10 & 10 & 5 & 0 & 0 \\ 
  ----- & ----- & ----- & ----- & ----- & ----- & ----- & ----- \\ 
  large & small &  &  &  &  &  &  \\ 
  0 & 0 &  &  &  &  &  &  \\ 
   \hline
\addlinespace
\multicolumn{8}{l}{\textbf{Mean shortfall from best}} \\
\midrule
  \hline
  \hline
EppsS & HHG & AD & Kuiper & small & large & Wassp1 & CvM \\ 
  5 & 8 & 16 & 25 & 28 & 33 & 45 & 53 \\ 
  ----- & ----- & ----- & ----- & ----- & ----- & ----- & ----- \\ 
  KS & ZA &  &  &  &  &  &  \\ 
  55 & 63 &  &  &  &  &  &  \\ 
   \hline
\addlinespace
\multicolumn{8}{l}{\textbf{Best selection(s)}} \\
\midrule
  \hline
  \hline
AD & HHG & Kuiper \\ 
   \hline
\bottomrule
\end{tabular}
\end{table}

\begin{table}[H]
\centering
\caption{Two-sample, discrete data, case studies of type skewness}
\label{tab:TDskewness}

\begin{tabular}{llllllll}
\toprule
\multicolumn{8}{l}{\textbf{Mean Power}} \\
\midrule
  \hline
  \hline
EppsS & HHG & small & Kuiper & AD & large & KS & CvM \\ 
  89 & 85 & 73 & 72 & 61 & 56 & 46 & 44 \\ 
  ----- & ----- & ----- & ----- & ----- & ----- & ----- & ----- \\ 
  Wassp1 & ZA &  &  &  &  &  &  \\ 
  43 & 27 &  &  &  &  &  &  \\ 
   \hline
\addlinespace
\multicolumn{8}{l}{\textbf{Percentage close to best}} \\
\midrule
  \hline
  \hline
EppsS & HHG & AD & Kuiper & CvM & Wassp1 & small & KS \\ 
  85 & 85 & 30 & 25 & 10 & 10 & 10 & 5 \\ 
  ----- & ----- & ----- & ----- & ----- & ----- & ----- & ----- \\ 
  ZA & large &  &  &  &  &  &  \\ 
  0 & 0 &  &  &  &  &  &  \\ 
   \hline
\addlinespace
\multicolumn{8}{l}{\textbf{Mean shortfall from best}} \\
\midrule
  \hline
  \hline
EppsS & HHG & small & Kuiper & AD & large & KS & CvM \\ 
  3 & 7 & 18 & 19 & 30 & 36 & 46 & 47 \\ 
  ----- & ----- & ----- & ----- & ----- & ----- & ----- & ----- \\ 
  Wassp1 & ZA &  &  &  &  &  &  \\ 
  48 & 65 &  &  &  &  &  &  \\ 
   \hline
\addlinespace
\multicolumn{8}{l}{\textbf{Best selection(s)}} \\
\midrule
  \hline
  \hline
HHG & Kuiper \\ 
   \hline
\bottomrule
\end{tabular}
\end{table}

\begin{table}[H]
\centering
\caption{Two-sample, discrete data, case studies of type kurtosis}
\label{tab:TDkurtosis}

\begin{tabular}{llllllll}
\toprule
\multicolumn{8}{l}{\textbf{Mean Power}} \\
\midrule
  \hline
  \hline
EppsS & HHG & Kuiper & large & small & AD & Wassp1 & KS \\ 
  89 & 86 & 66 & 65 & 61 & 59 & 38 & 35 \\ 
  ----- & ----- & ----- & ----- & ----- & ----- & ----- & ----- \\ 
  CvM & ZA &  &  &  &  &  &  \\ 
  33 & 21 &  &  &  &  &  &  \\ 
   \hline
\addlinespace
\multicolumn{8}{l}{\textbf{Percentage close to best}} \\
\midrule
  \hline
  \hline
EppsS & HHG & AD & large & Kuiper & Wassp1 & CvM & KS \\ 
  85 & 85 & 25 & 15 & 10 & 10 & 5 & 0 \\ 
  ----- & ----- & ----- & ----- & ----- & ----- & ----- & ----- \\ 
  ZA & small &  &  &  &  &  &  \\ 
  0 & 0 &  &  &  &  &  &  \\ 
   \hline
\addlinespace
\multicolumn{8}{l}{\textbf{Mean shortfall from best}} \\
\midrule
  \hline
  \hline
EppsS & HHG & Kuiper & large & small & AD & Wassp1 & KS \\ 
  4 & 7 & 27 & 28 & 31 & 34 & 55 & 58 \\ 
  ----- & ----- & ----- & ----- & ----- & ----- & ----- & ----- \\ 
  CvM & ZA &  &  &  &  &  &  \\ 
  60 & 72 &  &  &  &  &  &  \\ 
   \hline
\addlinespace
\multicolumn{8}{l}{\textbf{Best selection(s)}} \\
\midrule
  \hline
  \hline
AD & EppsS \\ 
   \hline
\bottomrule
\end{tabular}
\end{table}

\begin{table}[H]
\centering
\caption{Two-sample, discrete data, case studies of type mixed}
\label{tab:TDmixed}

\begin{tabular}{llllllll}
\toprule
\multicolumn{8}{l}{\textbf{Mean Power}} \\
\midrule
  \hline
  \hline
HHG & EppsS & Wassp1 & AD & Kuiper & small & CvM & KS \\ 
  79 & 74 & 64 & 62 & 62 & 59 & 50 & 48 \\ 
  ----- & ----- & ----- & ----- & ----- & ----- & ----- & ----- \\ 
  large & ZA &  &  &  &  &  &  \\ 
  46 & 38 &  &  &  &  &  &  \\ 
   \hline
\addlinespace
\multicolumn{8}{l}{\textbf{Percentage close to best}} \\
\midrule
  \hline
  \hline
HHG & EppsS & Kuiper & Wassp1 & CvM & AD & ZA & small \\ 
  80 & 55 & 40 & 40 & 30 & 30 & 20 & 10 \\ 
  ----- & ----- & ----- & ----- & ----- & ----- & ----- & ----- \\ 
  KS & large &  &  &  &  &  &  \\ 
  5 & 5 &  &  &  &  &  &  \\ 
   \hline
\addlinespace
\multicolumn{8}{l}{\textbf{Mean shortfall from best}} \\
\midrule
  \hline
  \hline
HHG & EppsS & Wassp1 & AD & Kuiper & small & CvM & KS \\ 
  5 & 9 & 20 & 21 & 22 & 25 & 34 & 35 \\ 
  ----- & ----- & ----- & ----- & ----- & ----- & ----- & ----- \\ 
  large & ZA &  &  &  &  &  &  \\ 
  38 & 46 &  &  &  &  &  &  \\ 
   \hline
\addlinespace
\multicolumn{8}{l}{\textbf{Best selection(s)}} \\
\midrule
  \hline
  \hline
CvM & HHG & small \\ 
  CvM & HHG & Kuiper \\ 
  AD & HHG & small \\ 
  HHG & small & Wassp1 \\ 
  AD & HHG & Kuiper \\ 
  HHG & Kuiper & Wassp1 \\ 
   \hline
\bottomrule
\end{tabular}
\end{table}

\section{Cross-validation of the best selections}\label{sec:loo}

The best selections were found from the same case studies on which they are judged. To see how well they carry over to new problems we use leave-one-out cross-validation: for each case study we find the best selection from all the other case studies of the same study and then check it on the case study that was left out. This needs no new simulations, only the stored powers. Table \ref{tblLOO} has the results. For comparison it also shows the regret of the same number of methods with the highest mean power, and of the single method with the highest mean power.

\begin{table}[H]
\centering
\begingroup\small
\begin{tabular}{p{2.8cm}rp{3.2cm}rrrrrr}
  \hline
Study & Cases & Selection (all cases) & Same & Covered & Regret & Top-k & Single & Worst \\ 
  \hline
Goodness-of-fit, continuous & 120 & AD, W, ZA, ZC, Wassp1, ES-s-L & 99 & 99 & 0.2 & 1.3 & 11.6 & 12 \\ 
  Goodness-of-fit, discrete & 120 & K, AD, Wassp1, NS-B, NS-A, BJ & 99 & 99 & 0.5 & 0.5 & 12.3 & 10 \\ 
  Two-sample, continuous & 100 & AD, ZC, EppsS, HHG & 99 & 99 & 0.7 & 2.4 & 9.3 & 10 \\ 
  Two-sample, discrete & 100 & Kuiper, AD, HHG & 100 & 100 & 2.1 & 0.8 & 5.7 & 9 \\ 
   \hline
\end{tabular}
\endgroup
\caption{Leave-one-out cross-validation of the best selections. Same: percentage of left out case studies for which the selection found from the other case studies is the same as the one from all case studies. Covered: percentage of left out case studies in which the selection contains a method within 90\% of the best. Regret: mean of the power of the best method minus the power of the best selected method, in percentage points. Top-k: the same for the same number of methods with the highest mean power. Single: the same for the single method with the highest mean power. Worst: largest regret of the selection.} 
\label{tblLOO}
\end{table}

The selections are very stable: in all four studies the selection found without one case study is the same as the one found from all case studies in at least $99\%$ of the cases, and the left out case study is covered (a selected method has a power within $90\%$ of the best) equally often. The mean loss of power by using the selection instead of the best method is less than one percentage point in the first three studies and $2.1$ percentage points for discrete data in the two-sample problem. There the three methods with the highest mean power (Epps-Singleton, HHG and Anderson-Darling) do slightly better ($0.8$ percentage points), the selection is made to have a method within $90\%$ of the best in every case study, not to have the smallest mean loss. Using the same number of methods with the highest mean power loses between $0.5$ and $2.4$ percentage points, and using a single method, even the one with the highest mean power, between $5.7$ and $12.3$ percentage points on average, which confirms that no single method can be recommended.

Note that our case studies are not independent, the case studies of the types mean to kurtosis use the same 20 distributions under the null hypothesis, and so leave-one-out cross-validation probably somewhat overstates how well the selections carry over.

\section{Continuous vs Discrete Methods}

\begin{figure}[H]
\centering
\includegraphics[width=0.8\textwidth]{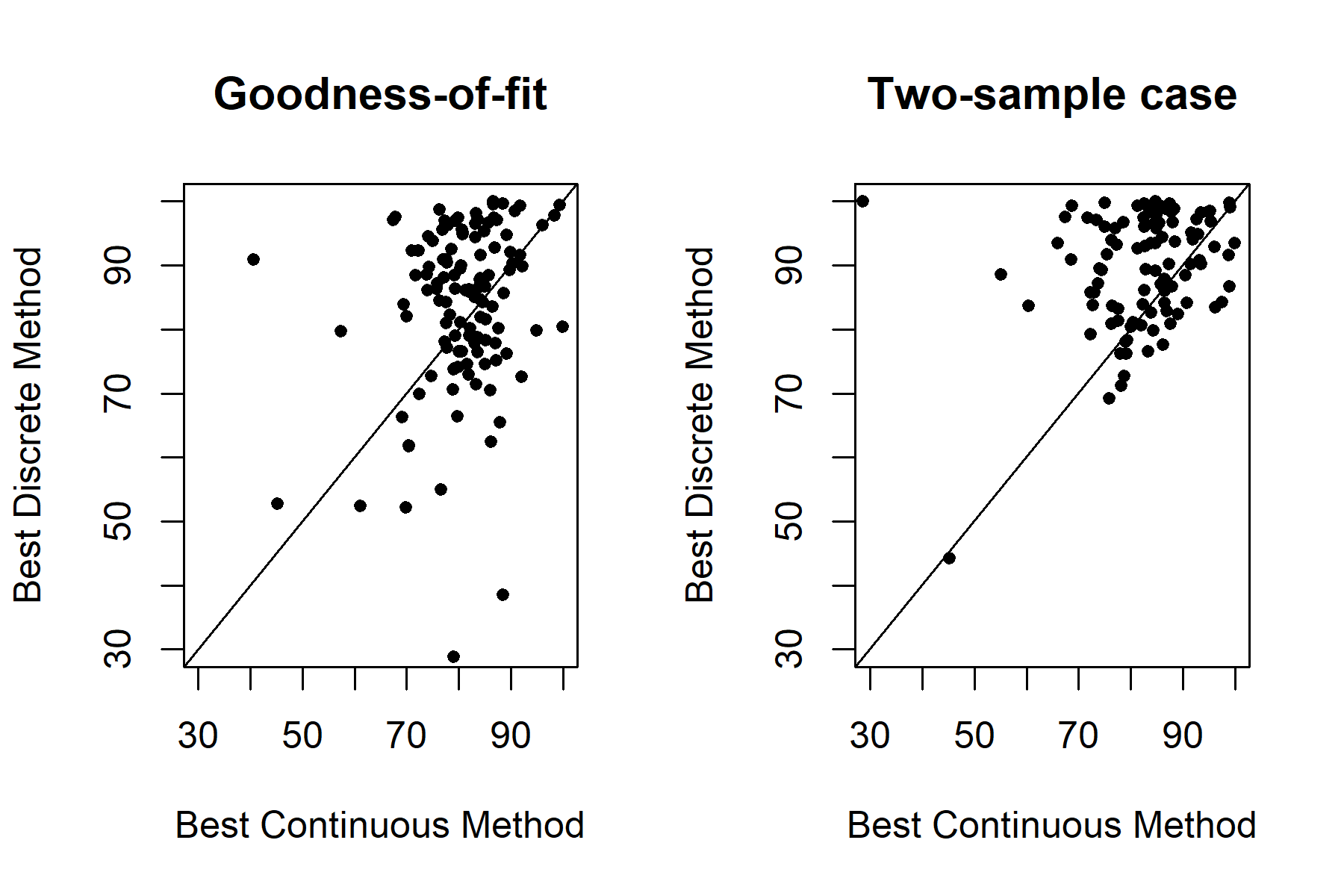}
\caption{Power of the best method for continuous data against the power of the best method for the discretized data.}
\label{fig:contdisc}
\end{figure}

In all the case studies we ran both continuous and discrete methods on the same data sets, after discretizing the continuous data set. This allows us to compare methods for both types of data. The result is shown in figure \ref{fig:contdisc}. In the goodness-of-fit problem the best method for continuous data has a higher power than the best method for discrete data in only $39\%$ of the case studies, and the median difference is about 3 percentage points in favor of the discrete methods. The main reason are the smooth tests and the Berk-Jones test, which are implemented in \emph{Rgof} only for discrete data. In the two-sample problem the best method for discrete data has a higher power in $69\%$ of the case studies (median difference about 5 percentage points), mostly because the Epps-Singleton and the HHG tests have a much higher mean power for the discretized data ($83\%$) than for the continuous data ($67\%$ and $63\%$).

Table \ref{tblCorCD} shows the correlation of the powers of the methods that exist for both types of data. For most methods based on the edf they are fairly high (between $53\%$ and $73\%$), in other words, a method that performs well for continuous data generally also performs well for the corresponding discretized data. They are lower for the Anderson-Darling test ($44\%$ and $33\%$), and close to zero for the chi-square tests in the two-sample problem. Here the bins for continuous data are found from the data, whereas for discrete data they are the given 50 values.

\begin{table}[H]
\centering
\begingroup\small
\begin{tabular}{rrrrrrrr}
  \hline
 & KS & Kuiper & CvM & AD & Wassp1 & Chisquare large & Chisquare small \\ 
  \hline
Goodness-of-fit & 72 & 64 & 73 & 44 & 64 & 52 & 55 \\ 
  Two-sample & 66 & 55 & 66 & 33 & 53 & 5 & 10 \\ 
   \hline
\end{tabular}
\endgroup
\caption{Correlation (in percent) over all case studies of the powers of a method for continuous data and for the discretized data.} 
\label{tblCorCD}
\end{table}

\section{True type I errors}\label{sec:typeI}

Tables \ref{tblNull1} to \ref{tblNull8} show the mean, median, smallest and largest true type I error of each method over the case studies of each study. In the goodness-of-fit problem with continuous data the mean type I errors of all methods are between $4.7\%$ and $5.5\%$, and the largest is $8.4\%$. The two-sample tests for continuous data are somewhat conservative, with mean type I errors between $3.2\%$ and $4.8\%$. In the goodness-of-fit and the two-sample problem with discrete data all mean type I errors are between $4.3\%$ and $5.4\%$. Over the four hybrid studies the mean type I errors are between $3.9\%$ and $5.3\%$, and the largest is $8.8\%$ (Kuiper, nMC=5, discrete data, normal-mix4). The largest stored type I errors occur in the goodness-of-fit problem with discrete data (Wasserstein in pareto.est and normal.cauchy.est, NS-B in gammamixture1.est); they are reproduced exactly with the seeds of the stored runs, but repeated runs with other random numbers show that they are not typical of these tests, see section \ref{sec:limitations}.

In an earlier version of these studies the methods based on the edf had a type I error of up to $18\%$ for discrete data in the case studies laplace1.est, laplace2.est and truncexp1.est, whereas the chi-square tests held the nominal level. The reason was the estimator, not the tests: the parameters were estimated by applying the estimator for continuous data (the median, the largest observation) to the midpoints of the bins. Such an estimator takes only a few distinct values, and the parametric bootstrap then fails to reproduce its distribution. The discrete versions of these case studies now use the grouped-data maximum likelihood estimator, which maximizes $\sum_j x_j\log\left(F(b_j)-F(b_{j-1})\right)$ over the bins $(b_{j-1},b_j]$, and all methods hold the nominal level. Users of the goodness-of-fit tests for binned data should therefore estimate the parameters from the binned data, not from the bin midpoints.

\input{tables/tblNull1.tex}

\input{tables/tblNull2.tex}

\input{tables/tblNull3.tex}

\input{tables/tblNull4.tex}

\input{tables/tblNull5.tex}

\input{tables/tblNull6.tex}

\input{tables/tblNull7.tex}

\input{tables/tblNull8.tex}

\section{Combining Several Tests}\label{sec:simtest}

The recommended selections can be used together, rejecting the null hypothesis if any of the selected tests does so. To keep the type I error at $\alpha$ the smallest p value then has to be adjusted, which can be done by simulation: the distribution of the smallest p value under the null hypothesis is found from simulated data sets and applied to the smallest p value of the actual data. This is implemented in the routines \emph{gof\textunderscore test\textunderscore adjusted\textunderscore pvalue} of \emph{Rgof} and \emph{twosample\textunderscore test\textunderscore adjusted\textunderscore pvalue} of \emph{R2sample}. We do not evaluate the power of such combined tests here.

\section{Limitations}\label{sec:limitations}

The conclusions of this paper are based on a collection of case studies chosen by the author, who is also the author of the two R packages that implement the methods. We have tried to include a wide variety of problems, but other kinds of alternatives might well favor other methods; the cross-validation in section \ref{sec:loo} checks the selections only within our collection. All studies use $\alpha=0.05$ and one sample size per case study; for the two-sample problem section \ref{sec:cases} shows that the rankings hardly change with the sample size, the level and the power, but we have not checked this for the goodness-of-fit problem. The parameter of each alternative was chosen so that the best method has a power of about $80\%$, which makes the mean powers depend on this choice.

The studies also revealed some weaknesses of individual methods. In a few case studies some methods could not be calculated, for example the Anderson-Darling test for the noncentral chi-square distribution and Anderson-Darling and Wasserstein for normal.t; in the summaries these were counted as a power of 0. In the goodness-of-fit problem with discrete data and parameter estimation the stored type I errors are $15\%$ for the Wasserstein test in the case study pareto.est, $11\%$ for the Wasserstein test in normal.cauchy.est and $9\%$ for NS-B in gammamixture1.est. These values are reproduced exactly with the seeds of the stored runs, but they are not typical: in ten repeated runs of each, with the same sample sizes and different random numbers, the type I error was between $1.2\%$ and $7.2\%$ (mean $4.6\%$) for the Wasserstein test in pareto.est, between $3.0\%$ and $6.3\%$ (mean $4.3\%$) for the Wasserstein test in normal.cauchy.est and between $2.3\%$ and $9.3\%$ (mean $5.5\%$) for NS-B in gammamixture1.est. So these tests hold the nominal level on average. In pareto.est and gammamixture1.est, however, the estimates vary about twice as much as is usual for the other methods (standard deviations of $2.1$ and $2.4$ percentage points, compared to about one point; because the critical value is itself estimated from $B=1000$ simulations, the standard deviation of a type I error estimate is about one percentage point rather than the binomial $\sqrt{0.05\cdot 0.95/1000}\approx 0.7$), because the critical value found from $B=1000$ simulations is unstable; for the Wasserstein test in pareto.est the reason is that the quantile function under the null hypothesis is unbounded, which makes the null distribution of the statistic very heavy-tailed. The largest values in the type I error tables are the most extreme of many such estimates, and the power estimates of these methods in these case studies are less reliable than the others.

\section{Conclusions}\label{sec:conclusions}

No single goodness-of-fit or two-sample test has good power against all alternatives: in most of our case studies in which the best method has a power of about $80\%$ some other method has a power below $30\%$. A small selection of methods, however, has a method close to the best in every case study, and the cross-validation shows that these selections carry over to case studies that were not used to find them. Using the selection instead of the best method loses at most $2.1$ percentage points of power on average, using the single method with the highest mean power between $5.7$ and $12.3$. For the two-sample problem with continuous data the rankings of the methods hardly depend on the sample size, the level or the power.

Which methods belong in the selection depends on the problem: Zhang's tests for continuous data without parameter estimation, Watson's test and the Anderson-Darling test when parameters are estimated, the data-driven smooth tests and the Berk-Jones test for discrete data in the goodness-of-fit problem, and the Epps-Singleton and HHG tests for discrete data in the two-sample problem. The complete recommendations are listed in section \ref{sec:highlights}. The tests of a selection can be combined into a single test with the adjusted p value of section \ref{sec:simtest}. For binned data the parameters should be estimated from the binned data, and hybrid tests should be used only when the cdf under the null hypothesis can not be computed. With the routines of \emph{Rgof} and \emph{R2sample} a new method can be compared with all of these on the same case studies.

\section{Detailed results}

In this section we show the detailed results of all the power studies. These are presented as follows. For each case study the methods are sorted according to their power, with the method with the highest power first. In the tables these are shown in dark red. Next come any method whose power is at least $90\%$ of the highest, and these are shown in light red. Any method with a power at least $75\%$ of best is in green and within $50\%$ in blue. Any method with a power less than $50\%$ of best is shown without color.

The actual powers, as well as the actual type I errors, are included as data sets in the R packages.

\setlength{\tabcolsep}{5pt}   
\renewcommand{\arraystretch}{0.75}

\subsection{Goodness-of-fit}

\subsubsection{Continuous data}

\input{tables/tblG.P.cont.mean.tex}

\input{tables/tblG.P.cont.variance.tex}

\input{tables/tblG.P.cont.skewness.tex}

\input{tables/tblG.P.cont.kurtosis.tex}

\input{tables/tblG.P.cont.mixed.tex}

\subsubsection{Discrete data}

\input{tables/tblG.P.disc.mean.tex}

\input{tables/tblG.P.disc.variance.tex}

\input{tables/tblG.P.disc.skewness.tex}

\input{tables/tblG.P.disc.kurtosis.tex}

\input{tables/tblG.P.disc.mixed.tex}

\subsection{Hybrid methods, nMC=1}

\subsubsection{Continuous data}

\input{tables/tblH1.P.cont.mean.tex}

\input{tables/tblH1.P.cont.variance.tex}

\input{tables/tblH1.P.cont.skewness.tex}

\input{tables/tblH1.P.cont.kurtosis.tex}

\input{tables/tblH1.P.cont.mixed.tex}

\subsubsection{Discrete data}

\input{tables/tblH1.P.disc.mean.tex}

\input{tables/tblH1.P.disc.variance.tex}

\input{tables/tblH1.P.disc.skewness.tex}

\input{tables/tblH1.P.disc.kurtosis.tex}

\input{tables/tblH1.P.disc.mixed.tex}

\subsection{Hybrid methods, nMC=5}

\subsubsection{Continuous data}

\input{tables/tblH5.P.cont.mean.tex}

\input{tables/tblH5.P.cont.variance.tex}

\input{tables/tblH5.P.cont.skewness.tex}

\input{tables/tblH5.P.cont.kurtosis.tex}

\input{tables/tblH5.P.cont.mixed.tex}

\subsubsection{Discrete data}

\input{tables/tblH5.P.disc.mean.tex}

\input{tables/tblH5.P.disc.variance.tex}

\input{tables/tblH5.P.disc.skewness.tex}

\input{tables/tblH5.P.disc.kurtosis.tex}

\input{tables/tblH5.P.disc.mixed.tex}

\subsection{Two-sample}

\subsubsection{Continuous data}

\input{tables/tblT.P.cont.mean.tex}

\input{tables/tblT.P.cont.variance.tex}

\input{tables/tblT.P.cont.skewness.tex}

\input{tables/tblT.P.cont.kurtosis.tex}

\input{tables/tblT.P.cont.mixed.tex}

\subsubsection{Discrete data}

\input{tables/tblT.P.disc.mean.tex}

\input{tables/tblT.P.disc.variance.tex}

\input{tables/tblT.P.disc.skewness.tex}

\input{tables/tblT.P.disc.kurtosis.tex}

\input{tables/tblT.P.disc.mixed.tex}

\bibliographystyle{apalike}
\bibliography{references}

\begin{thebibliography}{}

\bibitem[Anderson, 1962]{anderson1962}
Anderson, T.~W. (1962).
\newblock On the distribution of the two-sample {C}ramer-von { M}ises
  criterion.
\newblock {\em Annals of Mathematical Statistics}, 33(3):1148--1159.

\bibitem[Anderson and Darling, 1952]{anderson1952}
Anderson, T.~W. and Darling, D.~A. (1952).
\newblock Asymptotic theory of certain goodness-of-fit criteria based on
  stochastic processes.
\newblock {\em Annals of Mathematical Statistics}, 23:193--212.

\bibitem[Bellosta, 2011]{ad2011}
Bellosta, C.~G. (2011).
\newblock {\em Anderson-Darling GoF test with p-value calculation based on
  Marsaglia's 2004 paper Evaluating the Anderson-Darling Distribution}.
\newblock None.
\newblock R package version 0.3.

\bibitem[Berk and Jones, 1979]{berk1979}
Berk, R.~H. and Jones, D.~H. (1979).
\newblock Goodness-of-fit test statistics that dominate the {K}olmogorov
  statistics.
\newblock {\em Zeitschrift f{\"u}r Wahrscheinlichkeitstheorie und verwandte
  Gebiete}, 47:47--59.

\bibitem[Berkson, 1980]{berkson1980}
Berkson, J. (1980).
\newblock Minimum chi-square, not maximum likelihood.
\newblock {\em The Annals of Statistics}, 8(3):457--487.

\bibitem[Bickel and Doksum, 2015]{bickel2015}
Bickel, P.~J. and Doksum, K.~A. (2015).
\newblock {\em Mathematical Statistics: Basic Ideas and Selected Topics,
  Volumes I and II}.
\newblock CRC Press.

\bibitem[Casella and Berger, 2002]{casella2002}
Casella, G. and Berger, R. (2002).
\newblock {\em Statistical Inference}.
\newblock Duxbury Advanced Series in Statistics and Decision Sciences. Duxbury.

\bibitem[Cramer, 1928]{cramer1928}
Cramer, H. (1928).
\newblock On the composition of elementary errors.
\newblock {\em Scandinavian Actuarial Journal}, 1:13--74.

\bibitem[D'Agostino and Stephens, 1986]{agostini1986}
D'Agostino, R.~B. and Stephens, M.~A., editors (1986).
\newblock {\em Goodness-of-Fit Techniques}, volume~68 of {\em Statistics:
  Textbooks and Monographs}.
\newblock Marcel Dekker, New York.

\bibitem[Dolan and Mor{\'e}, 2002]{dolan2002}
Dolan, E.~D. and Mor{\'e}, J.~J. (2002).
\newblock Benchmarking optimization software with performance profiles.
\newblock {\em Mathematical Programming}, 91(2):201--213.

\bibitem[Dowd, 2022]{dowd2022}
Dowd, C. (2022).
\newblock {\em twosamples: Fast Permutation Based Two Sample Tests}.
\newblock None.
\newblock R package version 2.0.0.

\bibitem[Eddelbuettel et~al., 2024]{rcpp2024}
Eddelbuettel, D., Francois, R., Allaire, J., Ushey, K., Kou, Q., Russell, N.,
  Ucar, I., Bates, D., and Chambers, J. (2024).
\newblock {\em Rcpp: Seamless R and C++ Integration}.
\newblock R package version 1.0.12.

\bibitem[Epps and Singleton, 1986]{epps1986}
Epps, T.~W. and Singleton, K.~J. (1986).
\newblock An omnibus test for the two-sample problem using the empirical
  characteristic function.
\newblock {\em Journal of Statistical Computation and Simulation}, 26:177--203.

\bibitem[Fisher, 1922]{fisher1922}
Fisher, R.~A. (1922).
\newblock On the interpretation of chi square from contingency tables, and the
  calculation of p.
\newblock {\em Journal of the Royal Statistical Society}, 85:87--94.

\bibitem[Fisher, 1924]{fisher1924}
Fisher, R.~A. (1924).
\newblock The conditions under which chi square measures the discrepancy
  between observation and hypothesis.
\newblock {\em Journal of the Royal Statistical Society}, 87:442--450.

\bibitem[Goerg and Kaiser, 2009]{goerg2009}
Goerg, S.~J. and Kaiser, J. (2009).
\newblock Nonparametric testing of distributions - the {E}pps-{S}ingleton
  two-sample test using the empirical characteristic function.
\newblock {\em The Stata Journal}, 9(3):454--465.

\bibitem[Heller et~al., 2016]{heller2016}
Heller, R., Heller, Y., Kaufman, S., Brill, B., and Gorfine, M. (2016).
\newblock Consistent distribution-free {K}-sample and independence tests for
  univariate random variables.
\newblock {\em Journal of Machine Learning Research}, 17(29):1--54.

\bibitem[Kolmogorov, 1933]{kolmogorov1933}
Kolmogorov, A. (1933).
\newblock Sulla determinazione empirica di una legge di distribuzione.
\newblock {\em G. Ist. Ital. Attuari.}, 4:83--91.

\bibitem[Kuiper, 1960]{kuiper1960}
Kuiper, N.~H. (1960).
\newblock Tests concerning random points on a circle.
\newblock {\em Proceedings of the Koninklijke Nederlandse Akademie van
  Wetenschappen}, 63:38--47.

\bibitem[Ledwina, 1994]{ledwina1994}
Ledwina, T. (1994).
\newblock Data driven version of {N}eyman's smooth test of fit.
\newblock {\em J. Amer. Statist. Assoc.}, 89:1000--1005.

\bibitem[Lehmann, 1951]{lehmann1951}
Lehmann, E. (1951).
\newblock Consistency and unbiasedness of certain nonparametric tests.
\newblock {\em Ann. MAth. Statist.}, 22(1):165--179.

\bibitem[Neyman, 1937]{neyman1937}
Neyman, J. (1937).
\newblock Smooth test for goodness of fit.
\newblock {\em Skand. Aktuarietidskr}, 20:149--199.

\bibitem[Pettitt, 1976]{pettitt1976}
Pettitt, A.~N. (1976).
\newblock A two-sample anderson-darling rank statistic.
\newblock {\em Biometrika}, 63 No.1:161--168.

\bibitem[Rayner et~al., 2009]{raynor2009}
Rayner, J. C.~W., Thas, O., and Best, D.~J. (2009).
\newblock {\em Smooth Tests of Goodness of Fit: Using {R}}.
\newblock John Wiley \& Sons, Hoboken, NJ.

\bibitem[Rolke and Gutierrez~Gongora, 2021]{rolke2021}
Rolke, W. and Gutierrez~Gongora, C. (2021).
\newblock A chi-square goodness-of-fit test for continuous distributions
  against a known alternative.
\newblock {\em Computational Statistics}, 36:1885--1900.

\bibitem[Rosenblatt, 1952]{rosenblatt1952}
Rosenblatt, M. (1952).
\newblock Limit theorems associated with variants of the von mises statistic.
\newblock {\em Ann. Math. Statist.}, 23:617--623.

\bibitem[Smirnov, 1939]{smirnov1939}
Smirnov, N. (1939).
\newblock Estimate of deviation between empirical distribution functions in two
  independent samples.
\newblock {\em Bull. Moscow Univ.}, 2:3--16.

\bibitem[Thas, 2010]{thas2010}
Thas, O. (2010).
\newblock {\em Comparing Distributions}.
\newblock Springer Series in Statistics. Springer.

\bibitem[Vaserstein, 1969]{wasserstein1969}
Vaserstein, L.~N. (1969).
\newblock Markov processes over denumerable products of spaces, describing
  large systems of automata.
\newblock {\em Problemy Peredachi Informatsii}, 5(3):64--72.

\bibitem[von Mises, 1928]{mises1928}
von Mises, R. (1928).
\newblock {\em Wahrscheinlichkeit, Statistik und Wahrheit}.
\newblock Springer.

\bibitem[Watson, 1961]{watson1961}
Watson, G.~S. (1961).
\newblock Goodness-of-fit tests on a circle.
\newblock {\em Biometrika}, 48:109--114.

\bibitem[Zhang, 2002]{zhang2002}
Zhang, J. (2002).
\newblock Powerful goodness-of-fit tests based on likelihood ratio.
\newblock {\em Journal of the RSS (Series B)}, 64:281--294.

\bibitem[Zhang, 2006]{zhang2006}
Zhang, J. (2006).
\newblock Powerful two-sample tests based on the likelihood ratio.
\newblock {\em Technometrics}, 48:95--103.

\end{thebibliography}

\end{document}